\documentclass[11pt]{article}
\usepackage{amsmath, amssymb, amsfonts, amsthm}

\newtheorem{theorem}{Theorem}[section]

\newtheorem{lemma}[theorem]{Lemma}

\newcommand{\rd}{{\rm d}}
\newcommand{\be}{\begin{equation}}
\newcommand{\ee}{\end{equation}}
\newcommand{\bey}{\begin{eqnarray}}
\newcommand{\eey}{\end{eqnarray}}

\newcommand{\sfrac}[2]{{\textstyle \frac{#1}{#2}}}

\newcommand{\bp}{{\bf p}}

\newcommand{\bv}{{\bf v}}
\newcommand{\bu}{{\bf u}}

\newcommand{\bx}{{\bf x}}
\newcommand{\by}{{\bf y}}
\newcommand{\bz}{{\bf z}}

\newcommand{\balpha}{{\boldsymbol \alpha}}

\newcommand{\e}{\varepsilon}

\newcommand{\om}{{\omega}}

\newcommand{\bR}{{\mathbb R}}

\newcommand{\bN}{{\mathbb N}}
\newcommand{\bZ}{{\mathbb Z}}

\newcommand{\tr}{\mbox{Tr}}
\newcommand{\sgn}{\mbox{sgn}}

\newcommand{\wt}{\widetilde}
\newcommand{\wh}{\widehat}
\newcommand{\ov}{\overline}

\newcommand{\bxi}{{\boldsymbol \xi}}

\newcommand{\cS}{{\cal S}}

\newcommand{\bfeta}{{\boldsymbol \eta}}

\input epsf

\newcommand{\donothing}[1]{}

\begin{document}

\title{Nonlinear Hartree equation as the mean field limit \\ of weakly
coupled fermions}
\author{Alexander Elgart${}^1$,  L\'aszl\'o Erd\H os${}^2$\thanks{Partially
supported by NSF grant DMS-0200235. On leave from School of Mathematics,
GeorgiaTech, USA}
\\ Benjamin Schlein${}^1$ and
Horng-Tzer Yau${}^1$\thanks{Partially supported by NSF grant
DMS-0307295 and MacArthur Fellowship. On leave
from Courant Institute, New York University, USA} \\
\\
Department of Mathematics, Stanford University, CA-94305, USA${}^1$ \\ \\
Institute of Mathematics, University of Munich, \\
Theresienstr. 39, D-80333 Munich, Germany${}^2$
\\}

\date{November 18, 2003}

\maketitle

\begin{abstract}
We consider a system of $N$ weakly interacting fermions with a
real analytic pair interaction. We prove that for a general class
of initial data there exists a fixed time $T$ such that the
difference between the one particle density matrix of this system
and the solution of the non-linear Hartree equation is of order
$N^{-1}$ for any time $t\leq T$.
\end{abstract}



\section{Introduction}

The Hartree-Fock theory is a fundamental tool in atomic physics,
chemistry, plasma physics and many areas of quantum physics. It is
also an important numerical instrument to calculate atomic and
molecular structures. Despite numerous applications of the
Hartree-Fock theory, many basic theoretical questions remain
unsolved. One area where significant progress was made concerns
the ground state energy of large atoms and molecules.
Consider the  simple case of a neutral atom with nuclear
charge $Z$. It was first proved by Lieb and Simon \cite{LS1,LS2}
that the Hartree-Fock theory gives the correct asymptotic energy
to the leading order $Z^{7/3}$ as $Z\to \infty$. The next
important step came more than a decade later as Bach \cite{B}
proved that the error between the Hartree-Fock and the true atomic
energy is less than $Z^{5/3-\delta}$ for some small $\delta > 0$.
Similar result with a very different method was announced in
\cite{FS-1} and was proved in \cite{FS-2}.

The goal of this paper is to justify a time-dependent mean-field
theory for the evolution of interacting fermions under a weak pair
interaction with initial data localized in a cube of size of order
one. The last restriction actually provides the length scale of
the system. The interaction potential varies on the same length scale.
While one might want to add a background potential, we
shall keep the model simple to focus on the many-body interaction
effect. We work in $d=3$ dimensions, but our result holds in any
dimension. The Hamiltonian describing such a system is given by
\be
    H_N:= -  \frac  {\e^2} 2 \sum_{j=1}^N \Delta_{x_j} +
   \frac 1 N  \sum_{j,k}  U(x_j-x_k)
\label{def:HN}
\ee
acting on $\bigwedge_1^N L^2(\bR^3)$,
and the Schr\"odinger equation is given by
\be\label{Sch}
     i \e \partial_{t} \psi_t =  H_{N} \psi_t \;.
\ee
Here we have chosen the strength of the interaction between
fermions to be of order $1/N$. Examples of such systems with a
small coupling constant can be found in astrophysics and plasma
physics. For gravitating systems, the strength of the interaction
is dictated by the gravitational constant and thus the mean field
approximation is suitable. The Coulomb singularity, however, is
difficult to control. If one wishes to use (\ref{def:HN}) to model
the dynamics of white dwarfs, the kinetic energy has to be further
modified to be the relativistic one, according to the famous
observation by Chandrasekhar \cite{C}, see a rigorous account in
\cite{LY}. For the plasma physics application, the weak pair
potential models combined electron-electron and
electron-background interactions.

{F}rom now on we will fix a particular relation between $\e$ and
$N$, which is motivated by the following argument. We model a system of
$N$ fermions at energy comparable with the ground state energy
of the system. The potential
energy per particle is of order one. It is well-known that the
kinetic energy per particle of $N$ fermions, i.e.,
$-\frac{1}{2}\e^2 \Delta_{x_j}$, in a cube of size one scales like
${\e^2} N^{2/3}$ in the ground state.
In order to keep the kinetic energy per particle
of order one, we need to choose $\e=N^{-1/3}$, a convention we
shall use for the rest of this paper. With this choice, the
kinetic and the potential energy per particle in $H_N$ are
comparable. This is the basic physical criterion to obtain a
limiting dynamics (as $N\to\infty$) that captures the nonlinear
effect of the interaction. Notice that we have kept the free
evolution in the form of $i\e \partial_t\psi = -\frac{1}{2}\e^2
\Delta \psi$ so that the free evolution has a limit as $\e \to 0$.
The equation \eqref{Sch} is formally semiclassical with a
mean-field interaction potential at high density. Our choice of
scaling is the same as in \cite{NS} and \cite{S}. (The
interpretation of the origin of this scaling in \cite{NS} is
somewhat different.)

\bigskip

In order to take the limit $\e \to 0$,
we need to recast the Schr\"odinger equation
using the density matrix. For any wave function $\psi_{N,t}$,
define  the corresponding density matrix by $\gamma_{N, t} =
\pi_{\psi_{N, t}}$, where $\pi_{\psi} = |\psi \rangle \langle \psi
|$ is the orthogonal projection onto $\psi$. The kernel of
$\gamma_{N,t}$ is then given by
\begin{equation}
\gamma_{N,t} (\bx, \by) = \psi_{N,t} (\bx) \overline{\psi_{N,t}
(\by)}\; .
\end{equation}
The notation $\bx$ typically stands for
$\bx= (x_1, \ldots, x_N)$. Depending on the context sometimes it may
denote a shorter
vector of $x'$s.

We recall  that  a self adjoint operator $\gamma$ is called {\it
density matrix} if $0\le \gamma \le 1$ and $\tr \; \gamma =1$. If the
density matrix of the system is a one-dimensional projection then
we say the system is in  a pure state, otherwise it is in a mixed
state. The Schr\"odinger equation \eqref{Sch} is equivalent to the
Heisenberg equation for the density matrix:
\be \label{eq:schr}
    i\e \partial_t \gamma_{N, t} = [H_N, \gamma_{N, t}]\; ,
\qquad [A, B]= AB-BA\; .
\ee

The $n$-particle density matrix, $\gamma^{(n)}_{N,t}$, is defined
through its kernel
\be \label{nden}
    \gamma^{(n)}_{N, t} (x_1, ..,  x_n; y_1, ..,  y_n):=\int \rd
    x_{n+1} .. \rd x_N \; \gamma_{N, t}
    (x_1, ..,  x_n, x_{n+1} ,.., x_N; y_1, ..,
    y_n, x_{n+1},.., x_N)
\ee
for $1 \leq n \leq N$,  and  $\gamma^{(n)}_{N, t} :=0$
otherwise. Define the Wigner transform of the one particle density
matrix in the scale $\e$ by \be
    W_{N}^{(1)} (x; v):= \frac{1}{(2\pi)^{3}}
    \int  e^{-i v\cdot \eta } \gamma_N^{(1)}\Big(
        x + \e \frac{\eta}{2}, x - \e \frac{\eta}{2}\Big) \rd \eta \; .
\label{def:1wigner} \ee Recall the nonlinear Vlasov equation for a
phase space density $f$:
\be
    \partial_t  f_t(x, v)
    + v \cdot \nabla_{x}  f_t(x, v)
    = \nabla_x (U  \star \varrho_t )\cdot\nabla_v  f_t(x, v)\;,
\label{eq:vla} \ee
where
$$
    \varrho_t(x): =\int f_t(x, v)\rd v
$$
is  the configuration space density. It was proved by Narnhofer
and Sewell \cite{NS} that $W_{N}^{(1)}$  converges weakly to a
solution of the Vlasov equation \eqref{eq:vla} provided that the
Fourier transform of the potential is compactly supported, in
particular $U$ is real analytic. The regularity
assumption was substantially relaxed by Spohn \cite{S}.

Define
the  Hartree equation for the time dependent one-particle density
matrix $\om_t$ by
\be\label{H}
    i \e \partial_t\om_t = \Big[ -\frac{\e^2}{2}\Delta +
    U\star \varrho_t, \omega_t\Big]
\ee where $\varrho_t(x): = \om_t(x,x)$. Note that the Vlasov
equation (\ref{eq:vla}) is the semiclassical approximation of
(\ref{H}). One can extend this equation to the Hartree-Fock
equation by including the exchange term \be\label{HFt}
    i \e \partial_t\om_t = \Big[ -\frac{\e^2}{2}\Delta +
    U\star \varrho_t , \omega_t \Big]
    - \int \Big[ U(x-z)-U(y-z)\Big] \om_t(x, z)\om_t(z,y)\rd z \; .
\ee

Our main result proves that the Hartree equation correctly
describes  the evolution of the Schr\"odin\-ger equation
\eqref{eq:schr} up to order $O(\e)$. More precisely, it states
that for short semiclassical time the difference between the
Wigner transform $W_{N, t}^{(1)}$ of the solution to the
Schr\"odinger equation \eqref{eq:schr} and the Wigner transform of
the solution of the Hartree equation \eqref{H} is of order
$O(\e^3)$ provided that the potential $U$ is  real analytic. In
other words, all $\e^2$ corrections come from  the difference
between the Vlasov equation \eqref{eq:vla} and the Hartree
equation \eqref{H};  hence they are related to the accuracy of the
semiclassical approximation in the one-body theory. In particular
we show that all correlation effects are of order at most
$O(\e^3)$.

In fact, the main correlation effect, the
exchange term, is expected to be
order $\e^3$ for smooth potential and  $\e^2$ for the Coulomb
potential.
Our interpretation of the Hartree-Fock equation resembles
the theory concerning the
ground state energy for atoms where the Hartree-Fock
theory is proved to be correct
up to $\e^{2+\delta}$ smaller than the leading term \cite{B}. The
analyticity  condition and the short time  restriction of our result
is unsatisfactory; it nevertheless
shows what the correct formulation of the time-dependent Hartree
and Hartree-Fock theories should be.

In order to see the effects of the  exchange term, i.e. to show
that (\ref{HFt}) approximates the quantum dynamics even better
than (\ref{H}), we would need to consider $\e^3$ correction for
the smooth case or the $\e^2$ correction for the Coulomb
potential. Notice that our approach is perturbative and in
principle all $\e^3$ corrections, including the exchange terms,
can be calculated. However, there are other sources of $\e^3$
corrections (see the last two terms in \eqref{eq:iter} in
Section 4) which make the exchange correction
less prominent. This should be compared with the Coulomb case
where all $\e^2$ corrections are expected to be from the
semiclassical approximation to the Hartree equation and the
exchange terms.

In a recent paper Graffi et al.  \cite{P} proved the convergence
of the Wigner transform $W_{N}^{(1)}$ (of the solution of the
Heisenberg equation \eqref{eq:schr}) to the solution of the Vlasov
equation under the assumption that the initial wave function is
of the semiclassical form $\psi = Ae^{iS/\e}$. The  result also provided
error control and the proof is carried out by concise inequalities
as opposed to weak convergence method in \cite{NS} and \cite{S}.
The main restriction is the  initial wave functions to be
of the semiclassical form.
Although this type
of wave functions is suitable for bosons, fermionic wave functions
are antisymmetric and thus vanish frequently. Notice that in the
neighborhood of the zero set of the wave functions, the
semiclassical approximation is difficult to apply. In particular, one
naive attempt (for fermionic case) is to choose $S$ symmetric and
$A$ antisymmetric. But $\int |\nabla A|^2$ will be of order
$N^{3/5}$ and this violates a key assumption in this paper.

The recent   work of  Bardos et al. \cite{BGGM} considers the
equation
\be
  i \partial_t \psi_{N, t}  = \Bigg(- \alpha   \sum_{j=1}^N \Delta_{x_j}  +
   \frac 1 N  \sum_{j,k}  U(x_j-x_k)\Bigg)\psi_{N, t}
\label{BGMeq} \ee
with an arbitrary  $\alpha>0$ (we have put $\hbar=1$ which
is a constant of order one in this paper). In the
limit $N\to \infty$, it was proved that the difference between the
one-particle density matrix $\gamma_{N,t}^{(1)} =
|\psi_{N,t}\rangle\langle \psi_{N,t}|$ and the solution to  the
corresponding time-dependent Hartree-Fock equation vanishes in the
trace norm provided that the initial data is  a Slater determinant
(and some other assumptions). Notice that the time scale in
\eqref{BGMeq} is of order $\e=N^{-1/3}$ smaller than \eqref{Sch}.
Thus for initial data
considered in \cite{NS} \cite{S} and the present article, the one
particle dynamics  of \eqref{BGMeq} is governed by a free
evolution; the effect of the
interaction given by $U$ vanishes in the limit $N\to \infty$.
We shall make a more detailed comparison in Section 3.

Finally we comment on the method. Our approach is to based on the BBGKY
hierarchy
and iteration scheme. There are two major elements in the proof.
The first one is the control of error term. Since we work on the
BBGKY hierarchy for finite  $N$,
we need to control  the error term in the iteration scheme. Here
we used that the trace norm of the density matrix is preserved.
The second  observation concerns the combinatorics.
As usual, the BBGKY hierarchy will produce a $n!$ factor under iteration.
However, in the setting of this paper,
there are  extra sources of $n!$, for example, we will need to take high
moments
of the interaction:
\begin{equation}\label{eq:ass1.1}
\int |\hat{U} ( \xi )| |\xi|^m \rd\xi \sim C^m m!
\end{equation}
See \eqref{eq:ass1.1} for precise assumption. Since time ordered
integration provides only a $1/n!$, we will have to prove that the
combined effects of the factorials from all sources is just a
single $n!$. See the proof of Lemma 4.1 for details.

\bigskip

\emph{Acknowledgments. }We are grateful to Herbert Spohn for
useful discussions.

\section{Notations}
\setcounter{equation}{0}

We first fix the notations and recall some definitions. The
$n$-particle density matrix $\gamma^{(n)}_{N,t}$ is defined
through the equation \eqref{nden} and clearly satisfies the
following normalization \be
      \tr \; \gamma^{(n)}_{N, t}= 1\, .
\label{trace}
\ee
It is well-known that
the one particle density matrix satisfies the following operator
inequality (see \cite{L})
\be
    0\le  \gamma^{(1)}_{N, t} \le  \frac{1}{N} \; .
\label{trace1}
\ee
Therefore, we can write $\gamma^{(1)}_{N, t}$
as
$$
\gamma^{(1)}_{N, t}= \frac 1 N \sum_{j=1}^\infty a_j \pi_j
$$
where $\pi_j$ is the orthogonal projection onto $\varphi_j$, and
where $a_j \in [0,1]$ for all $j \in \bN$ with
$\sum_{j=1}^{\infty} a_j = 1$. Note
that in the definition of the $n$-particle density matrices
we followed the convention that the trace of the density matrices is
normalized.
In standard $N$-body theory an additional $N(N-1)\ldots (N-n+1)$ factor
would be present in \eqref{nden}.

\subsection{Wigner Transform}\label{sec:Wigner}

The Wigner transform of an $N$-body density matrix
$\gamma_N(\bx; \by)$ is defined by
\be
    w_N (\bx; \bv):= \frac{1}{(2\pi)^{3N}}
    \int  e^{-i\bv\cdot\by} \gamma_N\Big(
        \bx + \frac{\by}{2}, \bx - \frac{\by}{2}\Big) \rd\by \; .
\label{def:wigner} \ee {F}rom $\tr \; \gamma_N =1$ it follows that
\begin{equation}\label{eq:norm}
\int \rd\bx \rd\bv \, w_N (\bx ,\bv) = 1 .
\end{equation}
Since the velocities of the $N$ particles are of order $N^{1/3}=\e^{-1}$,
we rescale the Wigner transform $w_N$ so that its arguments be
typically of order one. Thus we defined the rescaled Wigner transform by
\begin{equation}
W_{N,\e} (\bx , \bv)= W_N (\bx , \bv) :=  \e^{-3N} w_N (\bx , \bv
/\e)= \frac{1}{(2\pi)^{3N}} \int \rd \by \; \gamma_N \Big( \bx +
\frac{\e \by}{2}, \bx - \frac{\e \by}{2} \Big) e^{-i\bv \cdot \by}
.
\end{equation}
The factor $\e^{-3N}$ guarantees that the normalization (\ref{eq:norm})
holds for
the rescaled Wigner transform $W_{N,\e} (\bx,\bv)$ as well.
The inverse transform is given by
$$
    \gamma_N (\bx, \by) = \int
    W_{N,\e} \Big( \frac{\bx + \by}{2}, \bu\Big)
    e^{i(\bx-\by)\cdot \bu/\e} \rd\bu .
$$
In particular, the particle density at the point $\bx$ is given by
$$
    \rho(\bx) : = \gamma_N (\bx, \bx) = \int W_{N,\e} ( \bx, \bu ) \rd\bu .
$$
In this paper, we are concerned
with the rescaled  Wigner transform only, so
we shall drop the adjective ``rescaled'' and
the $\e$ index from the notation. The rescaling parameter
$\e$ will always be related to the total number of particles as
$\e=N^{-1/3}$.

The time evolution of the Wigner transform $W_N (\bx, \bv)$ is
given by the Wigner equation
\begin{multline}\label{eq:rescwig}
    \partial_t W_N(t; \bx, \bv)
    + \sum_{j=1}^N v_j \cdot \nabla_{x_j} W_N(t; \bx, \bv)\\
    = -\frac{i \e^2}{(2\pi)^{3N}}
    \int \Bigg[ U\Big( \bx+\frac{\e\by}{2}\Big)-U\Big( \bx-\frac{ \e\by}{2}
    \Big)
    \Bigg] e^{i\by \cdot (\bu -\bv)} W_N(t; \bx, \bu)
    \rd\bu\rd\by ,
\end{multline}
which can be easily derived from the Heisenberg equation
(\ref{eq:schr}).

It is tempting to consider the Wigner transform as a probability
density on the phase space. The problem with this interpretation
is that $W_N (\bx,\bv)$ is not positive. In order to make the
Wigner transform positive we may take convolutions with
Gaussian distributions. We can define the \emph{Husimi function}
by
$$
    H_{N}^{\delta_1, \delta_2} : = W_N \star_x G^{(N)}_{\delta_1}
    \star_v G^{(N)}_{\delta_2}
$$
where $\star_x$ denotes the convolution in $x$-space and
$$
    G_\delta^{(N)} (\bz): = \frac{1}{(\pi\delta^2)^{3N/2}} \exp \Big( -
\frac{\bz^2}{\delta^2}
    \Big)
$$
is the centered Gaussian distribution in $3N$ dimensions with variance
$\delta$.
It is easy to check that $H_N^{\delta_1, \delta_2}\ge 0$ if
$\delta_1\delta_2\ge \e$. The Husimi function  is normalized
according to
$$
   \int  H_N^{\delta_1, \delta_2} (\bx, \bv)\, \rd \bx \rd \bv = \|\psi_N
   \|^2_2 = 1
$$
and thus can be considered as a probability density on the phase
space. The accuracy of the $H_N^{\delta_1, \delta_2}\ge 0$ is of
order $\delta_1$ for the space variables and $\delta_2$ for the
velocity variables in semiclassical units.

As a side remark, we recall that for $\delta_1:=\delta$,
$\delta_2:=\e \delta^{-1}$ the Husimi function is just the
standard \emph{Gaussian coherent state} at scale $\delta$:
$$
    H_{N}^{\delta, \e \delta^{-1}}(\bx, \bv) = C_{N ,\delta}(\bx, \bv)
    : = (2\pi \e)^{-3N} \langle \psi_N , \pi_{\bx, \bv}^\delta \psi_N
\rangle
$$
where $ \pi_{\bx, \bv}^\delta = | \phi_{\bx, \bv}^\delta \rangle
\langle \phi_{\bx, \bv}^\delta |$ is the orthogonal projection
onto the state
$$
    \phi_{\bx, \bv}^\delta (\bz) = \frac{1}{(\pi \delta^2)^{3N/4}}
    e^{i\bz\cdot \bv/\e} \exp{\Big( -\frac{(\bz-\bx)^2}{2\delta^2}\Big)}.
$$

The $k$-particle Wigner transform $W_N^{(k)}$ is defined to be the
Wigner transform of the $k$ particle density matrix
$\gamma_N^{(k)}$. Clearly, it can be viewed as the $k$-particle
marginal of $W_N$ since it satisfies
\be
\begin{split}
    W_N^{(k)}(x_1, \ldots x_k; v_1, \ldots v_k)
    = &\frac{1}{(2\pi)^{3k}} \int \rd \by \; \gamma_N^{(k)}
\Big(\bx+\frac{\e\by}{2}; \bx -\frac{\e\by}{2}\Big) \,
    e^{-i\bv \cdot \by} \\
    = &\int
    W_N(x_1, \ldots x_N; v_1, \ldots v_N) \rd x_{k+1}\ldots
    \rd x_{N} \rd v_{k+1} \ldots \rd v_N
\end{split}
\ee The $W_N^{(k)}$ are normalized as
$$
   \int W_N^{(k)}(\bx,\bv)  \rd \bx \rd \bv  =1.
$$
Notice this definition is consistent with the definition
\eqref{def:1wigner}.
We now give some examples of  $N$
particle wave functions.

\subsection{Some Examples}
\label{sec:ex}
One of the most important assumptions of our
results is that the $n$-particle Wigner transform $W_N^{(n)}$ of
the initial state is factorized in the limit $N \to \infty$. At
first glance this assertion might seem surprising, since we are
dealing with a system of fermions. In the following we present
some typical situation where this condition is indeed fulfilled
and we show a very atypical example where factorization is wrong.

\medskip

The standard examples of many-body fermionic states are the Slater
determinants and the quasifree states.

{\it 1. Slater determinants.}
For any orthonormal family $\{\varphi_j, j=1, \ldots,  N\}$ define
the determinant wave function
$$
\psi (\bx)  = \Big(\bigwedge_{j=1}^N \varphi_j \Big)(\bx): =
\frac{1}{\sqrt{N!}}
\sum_{\sigma \in S_N} \sgn (\sigma) \prod_{j=1}^N \varphi_j
(x_{\sigma_{j}})\; .
$$
The one particle density matrix
is given by
$$
    \gamma^{(1)}(x,x')= \frac{1}{N}\sum_{j=1}^N
\varphi_j(x)\ov{\varphi_j(x')} \; .
$$
The two-particle density matrix is
\begin{equation}\label{eq:gamma2}
\begin{split}
    \gamma^{(2)}(x,y,x',y') &= \frac{1}{2N (N-1)}\sum_{k,\ell=1}^N
     \Big[\varphi_k(x)\varphi_\ell(y)-\varphi_k(y)\varphi_\ell(x)\Big]
    \ov{\Big[\varphi_k(x')\varphi_\ell(y')-\varphi_k(y')
    \varphi_\ell(x')\Big]} \\ &=
    \frac N {N-1} \Big [ \gamma^{(1)}(x,x')\gamma^{(1)}(y,y')
    - \gamma^{(1)}(x,y')\gamma^{(1)}(y, x') \Big ] .
\end{split}
\end{equation}

{\it 2. Quasifree states.}  An $N$-particle state $\om$ is called
quasifree, if its $k$-particle density matrices
factorize by Wick theorem
$$
     \om^{(k)}( x_1, \ldots , x_k; y_1, \ldots , y_k)
     = \frac{N^k}{N(N-1)\ldots (N-k+1)}\; \mbox{det} ( \om^{(1)}(x_j, y_j)
)_{j=1,\ldots, k} \;.
$$
The unusual prefactor is present due to our choice of normalization.
In particular, quasifree states are
characterized by their one particle marginals. For example, Slater
determinants
are pure quasifree states.

The concept of quasifree state can be  generalized to grand
canonical states with an indefinite particle number. In
particular, any normalized density matrix  $\gamma$ on
$L^2(\bR^3)$, can be realized as a fermionic quantum state
whose one particle density matrix is $\gamma$.
The state can have expected particle number up to $1/\|\gamma\|$. In fact,
with $N := \int \rd x \tr \; (\omega a_x^\dagger a_x)$ we have $\gamma
(x,y) = N^{-1} \tr \; (\omega \, a_x^\dagger a_y )$, where
$a_x^\dagger$, $a_x$ are fermionic creation and annihilation operators.
Thus, for any
$\psi \in L^2 (\bR^3)$,
\begin{equation}
\langle \psi , \gamma \psi \rangle = \frac{1}{N} \tr \; (\omega
a_{\psi}^\dagger a_{\psi} ) = - \frac{1}{N} \tr \; (\omega a_{\psi}
a^\dagger_{\psi}) + \frac{\| \psi \|^2}{N} \leq \frac{\| \psi \|^2}{N},
\end{equation}
where the operators $a_{\psi}$ and $a^\dagger_{\psi}$ annihilate
and, respectively, create a fermion in the state $\psi$.

\bigskip

\emph{Example 1}: Let $\Omega: = [0, 2\pi]^3$ and consider the
states $\varphi_k(x)= (2\pi)^{-3/2} \,  e^{ikx}\chi(x\in \Omega)$
with $|k|\leq c N^{1/3}$, $k\in \bZ^3$. The number of states is
$O(N)$. We consider the pure state $\Psi = \bigwedge \varphi_k$ of
the $N$ particle system and we compute the marginals of its Wigner
transform. The one particle density matrix is given by
\be
\begin{split} \label{eq:g1}
\gamma^{(1)}(x,x')&=
\frac{1}{N}\sum_{k:|k|<cN^{1/3}} \varphi_k(x)\ov{\varphi_k(x')} \\
&= \frac{1}{N} \frac{\chi (x,x' \in \Omega)}{(2\pi)^3} \sum_{k:
|k| <c N^{1/3}} e^{ik(x-x')} \sim \chi(x, x'\in \Omega)
f((x-x')N^{1/3}) \end{split} \ee with some decaying function $f$,
such that $f(0)=1$. Its Wigner transform with rescaling parameter
$\e=N^{-1/3}$ is
$$
    W^{(1)}(x,v) =  \frac{\chi (x\in \Omega)}{(2\pi)^6 N}
    \sum_{k: |k|\leq c N^{1/3}}
    \int dy \; e^{i\e k y } e^{- i v y } \rightharpoonup
    \frac{1}{(2\pi)^3}\;\chi(x\in \Omega)\chi(|v|\leq c)
$$
when $N\to\infty$. Using (\ref{eq:gamma2}) the two particle
density matrix can be computed as well. Its Wigner transform is
given by
\begin{equation}
\begin{split}
W^{(2)} (\bx, \bv) = \frac{N}{(2\pi)^{6}(N-1)} \int \rd\by \;
e^{-i\bv \cdot \by} &\Big \{ \gamma^{(1)} \left(x_1 +\frac{\e
y_1}{2}, x_1 -\frac{\e y_1}{2}\right) \gamma^{(1)} \left(x_2
+\frac{\e y_2}{2}, x_2 -\frac{\e y_2}{2}\right) \\ &- \gamma^{(1)}
\left(x_1 +\frac{\e y_1}{2}, x_2 -\frac{\e y_2}{2}\right)
\gamma^{(1)} \left(x_2 +\frac{\e y_2}{2}, x_1 -\frac{\e
y_1}{2}\right) \Big\}
\end{split}
\end{equation}
Notice that the first term is $W^{(1)}(x_1, v_1)W^{(1)}(x_2, v_2)$
after neglecting the error $N/ (N-1) \cong 1$.

The second term is the so called exchange term and it vanishes as
$N\to\infty$. By (\ref{eq:g1}) this term can be written as
\begin{equation}\label{eq:exch}
W_{\text{ex}}^{(2)} (\bx, \bv) \cong \frac{\chi (x_1, x_2 \in
\Omega)}{(2\pi)^{12} N^2} \sum_{k,\ell} \int \rd\by \; e^{-i\bv
\cdot \by} e^{ik (x_1 - x_2) + ik \e (y_1 + y_2)} \, e^{i\ell (x_2
-x_1) + i\ell \e (y_1 + y_2)}
\end{equation}
Thus, for an arbitrary function $J (\bx,\bv)=J_1 (\bx) J_2 (\bv)$
we have
\begin{multline} \int \rd\bx \rd \bv J(\bx,\bv)
W^{(2)}_{\text{ex}} (\bx, \bv) \\= \frac{1}{(2\pi)^{12} N^2}
\sum_{k,\ell =1}^N  \; \int_{\Omega} \rd \bx J_1
(x_1,x_2) \; e^{i(k-\ell)(x_1 - x_2)} \int_{\Omega} \rd \by \, \hat{J}_2
(y_1,y_2) e^{i\e (k+\ell)(y_1 + y_2)}
\end{multline}
For any
smooth functions $J_1$, the $\bx$ integration is very small unless
$\ell\sim k$. Thus the order of the exchange term is $1/N$. Notice
that if we take $J(x, y) \sim |x-y|^{-1}$ then the exchange term
becomes of order $N^{-2/3}$, consistent with standard pictures
from semiclassical limits of atomic and molecular energies.
Indeed, since
$$
\int \rd\bx \; e^{ik x} \frac{1}{|x|}\sim
\frac{1}{|k|^2}\;,
$$
we obtain that in this case
$$
W_{\text{ex}}^{(2)} (\bx, \bv) \sim \frac{1}{N^2}\sum_{|k-\ell|
=1}^{cN}\frac{1}{(k-\ell)^2}\sim N^{-2/3}\;.
$$

Instead of choosing $\gamma^{(1)} (x,x')$ as in (\ref{eq:g1}), we
could also define
\begin{equation}
\gamma^{(1)} (x,x') = \frac{1}{N} \sum_{k \in \bZ^3} f(k)
\varphi_k (x) \ov{\varphi_{k} (x')}
\end{equation}
for an arbitrary distribution $f(k)$ with $0 \leq f(k) \leq 1$ for
all $k \in, \bZ^3$, and with $\sum_k f(k) =N$ so that
$\gamma^{(1)}(x,x')$ is a density matrix satisfying  the conditions
(\ref{trace}) and
\eqref{trace1}. Typical distributions are of the form $f(k) =
g(\e k)$ for some smooth function $g$. In this case the one
particle density matrix is supported within a distance of order
$\e$ from the diagonal, and the exchange term, analogously to
(\ref{eq:exch}), vanishes in the weak limit $N\to\infty$.

\bigskip

\emph{Example 2}: Let $\om$ be a smooth decaying function. We define
$$
\varphi_k (x) = \e^{-3/2}\omega \Big ( \frac {x- k} {\e} \Big )\; ,
$$
where $k$ runs over lattice sites with $|k|\le c/\e$, and
$\e=N^{-1/3}$, as always. In other words, $\varphi_k (x)$
represents a state localized inside a sphere of radius $\e$ around
the lattice site $k$. It is a straight-forward exercise to show
that the exchange term in the two-particle Wigner transform of
$\bigwedge \varphi_k$ is again of order $1/N$. Indeed, we obtain
in this case that
\begin{equation}
\begin{split}
\int \rd\bx J(\bx)
W^{(2)}_{\text{ex}} (\bx, \bv) = \frac{1}{(2\pi)^6 }
\sum_{k,\ell =1}^N  \int \rd\bx \rd\by &J(x_1,x_2) e^{-i\bv
\cdot \by} \, \omega \Big ( \frac {x_1- k}{\e} +\frac{y_1}{2} \Big
)\, \ov{\omega \Big ( \frac {x_2- k}{\e} - \frac{y_2}{2}\Big )}\\
&\times \omega \Big ( \frac {x_2- \ell}{\e} + \frac{ y_2}{2} \Big
)\,\ov{\omega \Big ( \frac {x_1- \ell}{\e} -\frac{y_1}{2} \Big )}
\\
= \frac{1}{(2\pi)^6 N^2}
\sum_{k,\ell =1}^N  \int \rd\bx \rd \by &J(\e x_1, \e x_2) e^{-i\bv
\cdot \by} \, \omega \Big ( x_1 - \frac {k}{\e} +\frac{y_1}{2}
\Big
)\, \ov{\omega \Big ( x_2 - \frac {k}{\e} - \frac{y_2}{2}\Big )}\\
&\times \omega \Big ( x_2 - \frac {\ell}{\e} + \frac{ y_2}{2} \Big
)\,\ov{\omega \Big ( x_1 - \frac {\ell}{\e} -\frac{y_1}{2} \Big )}
\\
\end{split}
\end{equation}
Clearly, for any smooth function $J(\bx)$, only terms with
$k=\ell$ give a considerable contribution to the sum; therefore
the right hand side of the above equation is bounded by $1/N$.

\bigskip

\emph{Example 3}: In the last two examples the kernel of the one particle
density matrix, $\gamma^{(1)}(x,y)$,
is concentrated on the diagonal $|x-y| \lesssim
N^{-1/3}$. Suppose now that we are given a one particle density
matrix $\gamma(x,y)$ on $[0,2\pi]^3\times [0,2\pi]^3$,
which satisfies $0\leq \gamma \leq \mbox{(const.)}/N$,  $\tr \; \gamma=1$
and it is supported near the diagonal. Let
\be
\tilde \gamma (x, y) :=
\beta\Big[ \gamma (x, y)+  \gamma (x+e, y)+  \gamma (x, y+e)+  \gamma (x+e,
y+ e)\Big]
\label{tildegamma}
\ee
with $e:=(0,0,2\pi)$.
We can choose the constant $\beta$ so that $\tr \; \tilde \gamma = 1$ and we
still have
$$
0\le \tilde \gamma  \le \mbox{(const.)} /N \; .
$$
Thus we constructed a  density matrix which is not concentrated on the
diagonal. The corresponding quasifree state can be constructed
by standard procedure.

This construction can be carried out on the level of the
wave functions as well. Let $\varphi_k$, $k\in \bZ^3$, $|k|\leq cN^{1/3}$
be $N$ orthonormal
one body wavefunctions supported the cube $[0,2\pi]^3$ as in Example 1. Then
$\gamma: = \frac{1}{N}\sum_k |\varphi_k\rangle\langle \varphi_k|$ is
supported near the diagonal.
Define
$$
\tilde \psi  =  \bigwedge_k \psi_k \; ,\qquad \mbox{with} \quad
\psi_k(x):=2^{-1/2} [\varphi_k (x)+
\varphi_k (x+ e)] \; .
$$
Then the one particle density matrix
is of the form $\tilde \gamma$ from (\ref{tildegamma}).
It is concetrated around three submanifolds $x=y$ and $x=y \pm e$
and not just along the  diagonal $|x-y| \lesssim N^{-1/3}$.
In particular, the exchange term is still of  order $1/N$.

\bigskip

The fact that the exchange terms are of order $1/N$ in all these
three examples tells us that there is a large class of initial
data for which $W_N^{(2)}$ is factorized in the weak limit $N\to
\infty$. Similar result can be obtained for any $n$-particle
function if $n$ is fixed: \be
   \lim_{N \to \infty}  W^{(n)}_N (x_1, \ldots , x_n, v_1, \ldots , v_n)
    = \prod_{j=1}^n W_N^{(1)} (x_j, v_j)\;,
\label{fact}
\ee
or, more precisely, $|\langle J , W^{(n)}_N - (W_N^{(1)})^{\otimes
n}\rangle| \leq c/N$,
where the constant depends on $n$ and on the
smooth test function $J (\bx , \bv)$.

\subsection{The Fourier Transform of $W_N (\bx,
\bv)$}\label{sec:mu}
Instead of working directly with the Wigner
function $W_N$ it is often more convenient to work with its
Fourier transform, which we define as
$$
    \mu_N (\bxi, \bfeta) : = \tr \; \gamma_N \, e^{-i(\e \bfeta \cdot
    \hat{\bp} + \bxi \cdot \hat{\bx})}
$$
where $\hat{\bx}$ and $\hat{\bp} = -i \nabla_{\bx}$ are the
position and momentum operators on $L^2 (\bR^{3N})$.
Narnhofer and Sewell defined the
same quantity with a somewhat different notation (see (3.2) in \cite{NS}):
the  $\bxi, \bfeta$ variables are interchanged
and the conjugate is considered.
Noting that
$$
\Big(e^{-i(\e \bfeta \cdot \hat\bp + \bxi \cdot \hat\bx)}\psi\Big)(\bx)  =
e^{i\frac{\e}{2} \bxi \cdot \bfeta} e^{-i\bxi \cdot \bx} \psi(\bx-
\e \bfeta),
$$
we have
$$
    \mu_N (\bxi, \bfeta)
    = \int e^{-i\bxi \cdot \bx} \gamma_N \Big( \bx- \frac{\e \bfeta}{2},
    \bx + \frac{\e \bfeta}{2}\Big)  \rd
    \bx = \int \rd \bx \rd \bv \; W_N (\bx,\bv)
    e^{-i\bxi \cdot \bx - i \bfeta \cdot \bv}
$$
and hence
$$
   W_N (\bx, \bv)= \frac{1}{(2\pi)^{6N}} \int \mu_N (\bxi, \bfeta)
   e^{i\bxi \cdot \bx+i\bfeta \cdot \bv}  \rd \bfeta \rd \bxi .
$$
Notice the operator norm of $e^{-i(\e \bfeta \cdot \hat\bp  + \bxi
\cdot \hat\bx)}$ is equal one. Since $\gamma_N$ is positive and $\tr
\; \gamma_N = 1$, we have \be
    |\mu_N (\bxi, \bfeta)| =
    |\tr \; \gamma_N e^{-i(\e \bp \cdot \bfeta  + \bxi \cdot \bx)}|
    \le 1 .
\label{mubound} \ee For any $J( \bx, \bv)$ with $\| \hat J
\|_{L_1(\rd \bxi, \rd \bfeta)}$ bounded, we have \be\label{wb}
|\langle J , W_N \rangle| = |\langle \hat J , \mu_N \rangle| \le
\| \hat J \|_{L_1(\rd \bxi, \rd \bfeta)} \, , \ee
where $\langle f , g \rangle :=\int \rd \bx \rd \bv
\ov{f(\bx,\bv)} g (\bx,\bv)$. Therefore, one can
always extract  weak limit points of Wigner transforms.

The time evolution of $\mu_N (\bxi, \bfeta)$ can be easily derived
from (\ref{eq:rescwig}):
\begin{equation}\label{eq:muNeq}
\begin{split}
\partial_t \mu_N (t,\bxi,\bfeta) = &\, \sum_{j=1}^N \xi_j \cdot
\nabla_{\eta_j} \mu_{N}
(t, \bxi,\bfeta) \\ &-  2 \e^2 \sum_{j < k}^N  \int \rd q \,
\hat{U} (q) \, \sin \left( \frac{\e}{2} q (\eta_j - \eta_k)\right)
\, \mu_N (t , \xi_1 , \dots, \xi_j - q , \dots , \xi_k + q , \dots ,\xi_N;
\bfeta),
\end{split}
\end{equation}
where we defined $\hat{U} (q) := (2\pi)^{-3} \int \rd x e^{-iqx}
U(x)$. We denote by $\mu_N^{(k)}$ the Fourier transform of the $k$
particle Wigner function $W_N^{(k)}$. Then we have
\begin{equation}
\begin{split}
\mu_N^{(k)} (\xi_1 \dots \xi_n;\eta_1 \dots \eta_n) &= \int \rd
\bx \rd \bv \; W_N^{(k)} (\bx, \bv) e^{-i\bx\cdot\bxi -i \bv \cdot
\bfeta} \\ &=  \mu_N (\xi_1,\dots , \xi_k , 0,\dots , 0; \eta_1,
\dots \eta_k, 0,\dots, 0),
\end{split}
\end{equation}
if $k\leq N$ and $\mu_N^{(k)} = 0$ otherwise. {F}rom
(\ref{mubound}) \be
      |\mu_N^{(k)} (\bxi, \bfeta)|\leq 1
\label{eq:sup} \ee is valid for all $k$.

\section{The BBGKY Hierarchy and the Main Result}
\label{sec:main}
\setcounter{equation}{0}
The family of marginals
$\{ W_N^{(n)} \}_{n =1,..,N}$ satisfies a hierarchy of equations,
usually called the BBGKY hierarchy, which can be derived from
(\ref{eq:muNeq}) (also using the symmetry of $W_N^{(n)}$):
\begin{equation}\label{eq:wighier}
\begin{split}
    \partial_t W_N^{(n)}(&t; x_1, \ldots ,x_n, v_1, \ldots ,v_n)
    + \sum_{j=1}^n v_j \cdot \nabla_{x_j} W_N^{(n)}(t;
    x_1, \ldots x_n, v_1, \ldots v_n) \\ = \, &-\frac{i\e^{3}}{(2\pi)^{3n}}
\, \sum_{1\leq j<k\leq n}
    \int \e^{-1} \Bigg[ U \Big( x_j+\frac{\e y_j}{2} -x_k-\frac{\e y_k}{2}
\Big)
    -U \Big( x_j-\frac{\e y_j}{2} -x_k+\frac{\e y_k}{2} \Big)
    \Bigg]\\&\times e^{i\sum_{j=1}^n y_j(u_j -v_j)}
    W_N^{(n)}(t; x_1,\ldots, x_n, u_1, \ldots, u_n) \rd u_1\rd y_1
    \ldots \rd u_n\rd y_n\\ &- \frac{i(1 - n\e^3)}{(2\pi)^{3n}} \sum_{j=1}^n
    \int \e^{-1} \Bigg[ U \Big( x_j+\frac{\e y_j}{2} -x_{n+1}
    \Big)
    -U \Big( x_j-\frac{\e y_j}{2} -x_{n+1} \Big)
    \Bigg]\, e^{i\sum_{j=1}^n y_j(u_j -v_j)} \\
    &\times W_N^{(n+1)}(t; x_1, \ldots, x_n, x_{n+1}, u_1, \ldots,
u_n,u_{n+1})
     \rd u_1\rd y_1
    \ldots \rd u_n\rd y_n \rd u_{n+1} \rd x_{n+1} .
\end{split}
\end{equation}
The main goal of this paper is to compare solutions of this
hierarchy of equation with tensor products of solutions of the one
particle Hartree equation (\ref{H}) which can be rewritten in
terms of the one particle Wigner transform as
\begin{equation}\label{HW}
\partial_t W_t (x,v) + v \cdot \nabla_x W_t (x,v) =
-\frac{i}{(2\pi)^3} \int \rd y \rd u \; \frac{1}{\e} \; \Big( (U
\star \rho_t) (x+\frac{\e y}{2}) - (U\star \rho_t)(x - \frac{\e
y}{2}) \Big) \, W_t(x,u) \, e^{iu\cdot y} .
\end{equation}
Let us denote by $\widetilde{W}^{(n)} (t , \bx, \bv)$ the $n$
particle Wigner transform constructed taking tensor products of
solutions of (\ref{HW}), that is
\begin{equation*}
\widetilde{W}^{(n)} (t , \bx, \bv) = \prod_{j=1}^n W_t (x_j, v_j).
\end{equation*}
Moreover we denote by
\begin{equation}
\begin{split}
H^{\delta_1 , \delta_2}_{\ell, N} (t, \bx ,\bv) &= \Big
(G_{\delta_1}^{(\ell)} \star_x G_{\delta_2}^{(\ell)} \star_v
W^{(\ell)}_N (t) \Big) (\bx , \bv) \qquad \text{and}\\
\widetilde{H}^{\delta_1 , \delta_2}_{\ell , N} (t, \bx , \bv)
&=\Big (G_{\delta_1}^{(\ell)} \star_x G_{\delta_2}^{(\ell)}
\star_v \widetilde{W}^{(\ell)} (t) \Big) (\bx , \bv)
\end{split}
\end{equation} the Husimi functions associated with the
solution $W_N^{(\ell)}(t)$ of (\ref{eq:wighier}) and with
$\widetilde{W}^{(\ell)} (t)$, respectively. Here we used, as in
Section \ref{sec:Wigner}, the notation
\begin{equation}
G_{\delta}^{(n)} (\bz) = \left( \frac{1}{\pi \delta^2
}\right)^{3n/2} \, e^{-\frac{\bz^2}{\delta^2}} \, .
\end{equation}

The main result of this paper can be stated as follows.
\begin{theorem}\label{thm} Let $U$ be a radial symmetric real valued
potential and
assume that there is a constant $\kappa_1$ so that
\begin{equation}\label{condi}
\| U \|_m = \int |\hat{U} (\xi)| |\xi|^m d\xi \leq \kappa_1^m m!
\end{equation}
for all $m \in \bN$. Suppose that, for $k\leq 2\log N$,
\begin{equation}\label{eq:ass3}
\Big| \Big\langle O^{(k)} , W_N^{(k)} (0) - \widetilde{W}^{(k)} (0)
\Big\rangle \Big| \leq \frac{1}{N} \sup_{\bx,\bv} |O^{(k)} (\bx,\bv)|.
\end{equation}
Then,  for any
fixed $\ell$, $\delta_1$ and $\delta_2$, we have
\begin{equation}\label{eq:mainre}
\limsup_{N \to\infty} \; \sup_{\bx , \bv} \Big| \left(
H^{\delta_1, \delta_2}_{\ell , N} - \widetilde{H}^{\delta_1 ,
\delta_2}_{\ell,N} \right) (t, \bx,\bv) \Big| \cdot N < \infty \,
\end{equation}
uniformly for all $t < \sfrac{1}{4} (\sqrt{1+ 1/(7\kappa_1^2)} - 1)$.
\end{theorem}
\emph{Remarks .} \begin{enumerate} \item[1)] Condition
(\ref{condi}) holds for bounded, real analytic functions $U(x)$.
The symmetry condition is physically natural. The proof can easily
be modified to include non-symmetric potentials as well. \item[2)]
It is clear from the proof (see Section \ref{sec:proof}) that the
theorem is still true, with $N$ in \eqref{eq:mainre} replaced by
$N^{-1+\kappa}$ with an arbitrary small $\kappa >0$, if we allow
$\delta_1$, $\delta_2$ and $\ell$ depend on $N$ as long the
conditions
\begin{equation}
\ell (N) = o(\sqrt{\log N})\;, \qquad [\delta_j (N)]^{-2} =
o(\sqrt{\log N}), \qquad j=1,2,
\end{equation}
are satisfied. \item[3)] It is also clear from the proof (see
Section \ref{sec:proof}) that the theorem is still true if we
replace the accuracy  $1/N$ both in \eqref{eq:ass3} and
\eqref{eq:mainre} by $N^{-\kappa}$ with  $0< \kappa\le 1$. It
follows  that if the exchange term $W_{\text{ex}}^{(n)}$ (see
Section \ref{sec:ex}) of the initial data is of order
$N^{-\kappa}$, then it remain of the same order for all
sufficiently small times.
\end{enumerate}

The proof of this theorem is given in Section \ref{sec:proof}
below: it is based on a perturbative expansion of solutions of the
BBGKY Hierarchy. For technical reasons, instead of expanding
solutions of (\ref{eq:wighier}), it turns out to be more
convenient to work with the Fourier transforms $\mu_N^{(\ell)}
(\bxi,\bfeta)$ of the $W_N^{(\ell)} (\bx,\bv)$ (see Section
\ref{sec:mu} for the definition of $\mu_N^{(\ell)}$). Eq.
(\ref{eq:wighier}) is equivalent to the following hierarchy of
equations for the marginals $\mu_N^{(n)}$:
\begin{equation}\label{eq:BBGKYmu}
\begin{split}
&\partial_t \mu_N^{(n)}(t, \bxi, \bfeta)
    = \, \sum_{j=1}^n \xi_j \cdot \nabla_{\eta_j} \mu_N^{(n)}(t, \bxi,
\bfeta) \\
    &- 2 \e^2 \sum_{1\leq j < k \leq n}
    \int \rd q \hat{U}(q) \, \sin
    \left(\frac{\e\ q\cdot(\eta_j - \eta_k)}{2}\right)\,
    \mu_N^{(n)}(t, \xi_1, \dots, \xi_j - q,\dots \xi_k + q, \dots, \xi_n,
\bfeta) \\
    &- (1-n\e^3)\sum_{j=1}^n
    \int \rd q \hat{U}(q) \, \frac{2}{\e} \sin \left(
    \frac{\e q\cdot\eta_j}{2}\right)
    \mu_N^{(n+1)}(t, \xi_1, \ldots \xi_j - q ,\dots \xi_n, q ,
    \bfeta, 0)
\end{split}
\end{equation}

\subsection{Vlasov hiearchy}

The classical Vlasov hierarchy is the semiclassical approximation
of the BBGKY hierarchy. It is obtained from (\ref{eq:wighier}) by
formally setting $\e \to 0$ and approximate the potential
difference by gradient. Using \be
     U \Big( x_j+\frac{\e y_j}{2} -x_{n+1}
    \Big)
    -U \Big( x_j-\frac{\e y_j}{2} -x_{n+1} \Big)
    = \nabla U  (x_j-x_{n+1})  \e y_j + O(\e^2)
\label{eq:taylor} \ee and
$$
    iy_j e^{iy_j(v_j-u_j)} = -\nabla_{u_j}  e^{iy_j(v_j-u_j)},
$$
we can perform an integration by parts, then integrate out $\rd
u_1\rd y_1 \ldots \rd u_n\rd y_n$ to collect delta functions
$\prod_{1}^n \delta(u_j-v_j)$. Neglecting lower order terms,  we
obtain formally
\be
    \partial_t \wt W^{(n)}(t; x_1, \ldots x_n, v_1, \ldots v_n)
    + \sum_{j=1}^n v_j \cdot \nabla_{x_j} \wt W^{(n)}(t;
    x_1, \ldots ,x_n, v_1, \ldots ,v_n)
\label{eq:vlahier} \ee
$$
    = \sum_{j=1}^n
    \int\nabla U   (x_j-x_{n+1})
    \nabla_{v_j} \wt W^{(n+1)}(t; x_1, \ldots ,x_n, x_{n+1},
    v_1, \ldots ,v_n, u_{n+1})
    \rd u_{n+1} \rd x_{n+1}
$$
for the weak limit
\be
    \wt W^{(n)}(t, \bx, \bv): = \lim_{N\to\infty} W^{(n)}_{N} (t, \bx, \bv)
\; .
\label{eq:lim} \ee The main result of \cite{NS} and \cite{S}
proves that this limit exists, it solves the Vlasov hierarchy
(\ref{eq:vlahier}) and the solution is unique. Therefore $\wt
W^{(n)} =  w_t^{\otimes (n)}$ where $w_t$ satisfies the Vlasov
equation \eqref{eq:vla}.

\subsection{Other Scalings}

Theorem \ref{thm} identifies the limit dynamics of the Wigner transform
at scale $\e=N^{-1/3}$. One may define the Wigner transform at a different
scale $\nu$
by
\be
    W_{N, \nu}^{(1)} (x; v):= \frac{1}{(2\pi)^{3}}
    \int  e^{-i v\cdot y } \gamma_N^{(1)}\Big(
        x + \nu \frac{y}{2}, x - \nu \frac{y}{2}\Big) \rd y \; .
\label{def:1wignernu}
\ee

The following lemma
shows, however,  that under a natural energy conditions, $W_{N, \nu}^{(1)}$
cannot converge to a non-trivial
function unless $\nu\sim \e$. Similar statement is true
for higher order marginals. It justifies our choice of scaling in Section
\ref{sec:Wigner} in order to derive a dynamics for a nondegenerate limiting
distribution.

\begin{lemma}\label{lemma:wigner} Let $\e:=N^{-1/3}$.
\begin{itemize}
\item[(i)] Suppose the kinetic energy of a state $\Psi$ is
comparable with the ground state kinetic energy of $N$ fermions in
a box of size one, i.e. \be
      \Big\langle \Psi, \Big(\sum_{j=1}^N -\Delta_{x_j}\Big) \Psi\Big\rangle
      \leq C_1 N^{5/3} \; .
\label{eq:kin}
\ee
Let  $O\in \cS(\bR^3_x\times\bR^3_v )$ be a Schwarz function with support in
$\{ |v|\ge \lambda\}$ for some $\lambda>0$.
Then
\be
     |\langle O, H_{N, \nu} \rangle| \leq \Big[
     C_1 \Big(\frac{\nu}{\lambda\e}\Big)^2 + O(\nu\lambda^{-2})\Big]
     \| O\|_\infty \;,
\label{eq:away} \ee with $H_{N, \nu}:=W_{N, \nu}^{(1)}\star_x
G_{\sqrt{\nu}}\star_v G_{\sqrt{\nu}}$, in particular, the  one
particle Husimi function of $\Psi$ at scale $\nu$ vanishes outside
of the $\{ v=0\}$ hyperplane if $\nu\ll \e$.
\item[(ii)] Suppose
that the average mean square displacement of $\Psi$ is of order
one, i.e. \be
      \Big\langle \Psi, \frac{1}{N}\Big(\sum_{j=1}^N x_j^2\Big)
\Psi\Big\rangle
      \leq C_2 \; .
\label{eq:sq}
\ee
Let  $O\in \cS(\bR^3_x\times\bR^3_v )$ be a Schwarz function with support in
$\{ |v|\le \lambda\}$ for some $\lambda>0$.
Then
\be
     |\langle O, H_{N, \nu} \rangle| \leq \mbox{(const.)}
     C_2 \| O \|_\infty \Big( \frac{\lambda\e}{\nu}\Big)^{6/5}
\label{eq:in} \ee with a universal constant.
\end{itemize}
\end{lemma}

\begin{proof}
(i) Since $|O(x,v)|\leq \| O\|_\infty v^2\lambda^{-2}$ and
the Husimi function is positive, we obtain
$$
     |\langle O,
     H_{N, \nu}\rangle|\leq \frac{\| O\|_\infty }{\lambda^2}
     \langle v^2,  H_{N, \nu}\rangle = \frac{\| O\|_\infty }{\lambda^2}
     \int u^2 G_{\sqrt{\nu}}(v-u)
     \varrho_\nu (v) \rd v \rd u = \frac{\| O\|_\infty }{\lambda^2}
     \int (v^2 + c\nu)\varrho_\nu (v) \rd v \;,
$$
where $\varrho_\nu(v): =  \int W(x,v) \rd x$ is the momentum
distribution and $c$ is a universal constant.
Since $\int \varrho_\nu(v) \rd v =1$ and by \eqref{eq:kin}
$$
     \int v^2\varrho_\nu(v)\rd v = \nu^2 \; \tr (-\Delta)\gamma \leq C_1\;
     \nu^2 N^{2/3} = C_1 (\nu/\e)^2\;,
$$
we obtain (\ref{eq:away}).

(ii) We apply the Lieb-Thirring inequality \cite{LT} in the
Fourier space \be
        \int \varrho(v)^{5/3}\rd v \leq  \mbox{(const.)}
        \Big\langle \Psi, \Big(\sum_{j=1}^N x_j^2\Big) \Psi\Big\rangle  \;
\label{eq:LT}
\ee
with a universal constant,
where $\varrho(v) = N\int |\wh\Psi (v, v_2, \ldots , v_N)|^2\rd v_2\ldots
\rd v_N$
is the one particle momentum density of the antisymmetric function $\Psi$.
After rescaling we obtain $\varrho_\nu(v)= (\e/\nu)^3 \varrho(v/\nu)$, hence
$\int \varrho_\nu^{5/3} \leq \mbox{(const.)} C_2  (\e/\nu)^2$
from \eqref{eq:sq} and \eqref{eq:LT}. Therefore
\be
\begin{split}
      |\langle O, H_{N, \nu} \rangle| &\leq \| O \|_\infty \int \rd x \rd v
\;
      \chi( |v| \leq \lambda) H_{N,\nu}(x,v) \\
      &= \| O \|_\infty \int \rd u \rd v \;
      \chi( |u| \leq \lambda) G_{\sqrt{\nu}}(v-u)
     \varrho_\nu (v)  \\
     &\leq  \mbox{(const.)}\| O \|_\infty \| \chi( |\cdot | \leq
\lambda)\|_{5/2}
     \| G_{\sqrt{\nu}}\|_1 \|\varrho_\nu\|_{5/3} \\
     &\leq  \mbox{(const.)}  C_2 \| O \|_\infty \Big(
\frac{\lambda\e}{\nu}\Big)^{6/5}
\end{split}
\ee
by Young's inequality.
\end{proof}

This lemma shows that the weak limits of $W_{N,\nu}^{(1)}$ are
zero if $\nu\gg \e$, in particular if the Wigner transform is
unscaled, $\nu=1$.  It may, nevertheless, be reasonable to investigate
how well Hartree or Hartree-Fock evolutions approximate the true dynamics
compared to the actual size of $W$ in a different topology.

\bigskip
Bardos et al. \cite{BGGM} have recently studied the equation
\eqref{BGMeq} and showed that the Hartree-Fock equation
approximates the dynamics in the trace norm. In order to study
\eqref{BGMeq} one first needs to choose the parameter $\alpha$.
Denote by $H_{N,\alpha} = -\alpha \Delta_N + (1/N) \sum_{i<j}
U(x_i -x_j)$ the Hamiltonian corresponding to \eqref{BGMeq}, and
consider an initial state $\gamma_{N,0}$. Here we assume the two
body potential $U$ to have bounded derivative. Let $\gamma_{N,t}$ be
the time evolution of $\gamma_{N,0}$. We
are interested in an
estimate for the mean squared distance between two particles at an
arbitrary fixed time $t$. Define the quantities
\begin{equation*}
\begin{split}
u_t &:= \big[ \tr \; \gamma_{N,t} \, (x_1 - x_2)^2  \big]^{1/2},\\
v_t &:= \big[\tr \; \gamma_{N,t} \, (p_1 - p_2)^2 \big]^{1/2} \; ,
\end{split}
\end{equation*}
where  $ p_j := -i\nabla_{x_i}$.
For typical interacting initial states
the mean square distance between the particles is of order one, $u_0= O(1)$,
the kinetic energy per particle is of order $N^{2/3}$, due to
Fermi statistics, therefore $v_0\leq (const.)N^{1/3}$. The next
lemma shows that $v_0$ is exactly of order  $N^{1/3}$
for any fermionic state localized within an order one distance
from the center of mass; in particular there cannot be
strong velocity correlation between the particles. Then in Lemma \ref{lemma:mourre}
we show how to use the lower bound on $v_0$ to give a lower bound on
the mean square displacement $u_t^2$.

\begin{lemma}\label{lemma:lt}
Let $\gamma$ be a fermionic $N$-particle density matrix, $\tr \; \gamma=1$,
 satisfying
\be
       \tr \; \Big[ \gamma\; \frac{1}{N}\sum_{j=1}^N (x_j - \bar X)^2  \Big] \leq K \; ,
       \qquad  \bar X: = \frac{1}{N} \sum_{j=1}^N x_j \;.
\label{eq:CM}
\ee
Then
$$
     \tr \;   \gamma \,\,  (p_1-p_2)^2 \ge (const.)N^{2/3}
$$
 with a positive constant depending on $K$.
\end{lemma}

{\it Remark.} By the symmetry of $\gamma$ and a Schwarz inequality
\be
\begin{split}
      \tr \; \Big[ \gamma\; \frac{1}{N}\sum_{j=1}^N (x_j - X)^2  \Big]
    & =  \tr \; \gamma \Big( \frac{1}{N}\sum_{j=1}^N (x_1-x_j)\Big)^2 \\
      &\leq \frac{N-1}{N} \; \tr \;  \gamma \, (x_1-x_2)^2\; ,
\end{split}
\ee
so the condition (\ref{eq:CM}) is satisfied if
 $\tr \;  \gamma \, (x_1-x_2)^2 \leq K$.

\begin{lemma}\label{lemma:mourre}
Let $C:=\| \nabla U \|_\infty$ and let $u_0$, $v_0$ be the initial mean squared
distance between two particles in position and momentum space, respectively.
 Then for any $0\leq t \leq  v_0/(8C)$
\be
        u_t^2 \ge u_0^2+ \alpha^2\,  v_0^2 t^2 - (const.)\alpha t
        \Big(u_0 v_0 + u_0 t + \alpha v_0 t^2 \Big)\; ,
\label{eq:lowu} \ee
where the constant depends only on $C$.
\end{lemma}
The proofs of these lemmas are deferred to the Appendix.

\bigskip

According to Lemma \ref{lemma:lt} and the subsequent remark, if the initial inter-particle distance $u_0$  is
of order one, then $v_0\ge (const.)N^{1/3}$. In this case Lemma \ref{lemma:mourre}
shows that if we want $u_t$ to remain of order one for $t> 0$
uniformly as $N\to \infty$, then we have to assume that $\alpha
= O(\e) =O(N^{-1/3})$. Otherwise the interaction between the particles
typically vanishes as $U(x_1-x_2)\to 0$ for $|x_1-x_2|\to\infty$.

When $\alpha = \e$,  we can rewrite  the Schr\"odinger equation
(\ref{BGMeq}) as
\be
  i \e \partial_t \psi_{N, t}  = \Bigg(- \e^2  \sum_{j=1}^N \Delta_{x_j}  +
   \frac \e N  \sum_{j,k}  U(x_j-x_k)\Bigg)\psi_{N, t} .
\label{BGMeq2}
\ee
This equation is the same as (\ref{Sch}) except the extra $\e$ factor
in front of the interaction. Since (\ref{Sch}) converges to the Vlasov
equation, (\ref{BGMeq2}) converges to a free evolution.

Although some of these conclusions are partly
 based on initial data considered in \cite{NS},
\cite{S} or Section \ref{sec:ex},  this behavior is expected for a general
reasonable interacting physical system.  While one may
be able to consider  some initial data  so that the one particle
density matrix $ \gamma^{(1)}_{N, t}$ (for the dynamics \eqref{BGMeq}) is
not given by a free evolution in the $N\to\infty$ limit,
 we do not know if there is a natural class of such initial data.

\section{Proof of the Main Result}
\label{sec:proof}
\setcounter{equation}{0}
As explained in Section
\ref{sec:main} the proof of our main result, Theorem \ref{thm}, is
based on a perturbative expansion of solutions $\mu_N^{(\ell)} (t,
\bxi, \bfeta)$ of the BBGKY hierarchy in the form
(\ref{eq:BBGKYmu}). We will compare $\mu_N^{(\ell)} (t, \bxi,
\bfeta)$ with tensor products of a solution of the Hartree
equation (\ref{HW}), which, after Fourier transform can be written
in the form
\begin{equation}\label{eq:hartree}
\partial_t \mu_t (\xi,\eta) =  \, \xi \cdot \nabla_{\eta} \mu_t (\xi ,
\eta) - \int \rd q \; \hat{U} (q) \frac{2}{\e} \sin \left(
\frac{\e q\eta}{2}\right) \mu_t (\xi-q, \eta) \mu_t (q ,0)
\end{equation}
with a given initial condition $\mu_0$.
In the following we will use the notation
\be
\tilde{\mu}^{(\ell)}_t (\bxi, \bfeta) = \prod_{j=1}^{\ell} \mu_t (\xi_j,
\eta_j)
\label{eq:Hprod}
\ee
for $\ell$-particle tensor products of a solution $\mu_t$ of
(\ref{eq:hartree}). We remark that global existence, uniqueness
and regularity of the solution of (\ref{eq:hartree}) have been
established in \cite{GV-1, GV-2}.

For any $n$-particle observable $O^{(n)} (\bxi,\bfeta)$, with
$\bxi =(\xi_1, \ldots , \xi_n)$, $\bfeta=(\eta_1, \ldots ,
\eta_n)$, we define the norms
\begin{equation}
\| O^{(n)} \|_{\balpha} = \int \rd\bxi \rd\bfeta \; |O^{(n)}
(\bxi,\bfeta)| \prod_{j=1}^n (|\xi_j| + |\eta_j|)^{\alpha_j}
\end{equation}
for $\balpha = (\alpha_1 , \dots ,\alpha_n) \in \bN^n$. Moreover
we use the notation
\begin{equation}
\langle O^{(n)} , \mu^{(n)} \rangle = \int \rd\bxi \rd\bfeta \;
\ov{O}^{(n)} (\bxi ,\bfeta) \; \mu^{(n)}(\bxi,\bfeta).
\end{equation}

The following lemma is the main ingredient in the proof of Theorem
\ref{thm}.

\begin{lemma}\label{lm:mu}
Assume there exists a constant $\kappa_1$ so that
\begin{equation}\label{eq:ass1}
\| U \|_m = \int |\hat{U} ( \xi )| |\xi|^m \rd\xi \leq \kappa_1^m
m!
\end{equation}
for all $m\in \bN$. Fix positive integers $\ell , n$ and suppose that, for
all $k \leq (n+\ell)$,
\begin{equation}\label{eq:ass}
    \Big|\langle O, \mu_N^{(k)}(0)-\tilde{\mu}_0^{(k)} \rangle \Big|
    \leq \frac{1}{N} \| O\|_0.
\end{equation}
Consider an observable $O^{(\ell)}$ with
\begin{equation}\label{eq:ass2}
\| O^{(\ell)} \|_{\balpha} \leq C_0^{\ell} \, \kappa_2^{|\balpha|}
\alpha_1! \dots \alpha_{\ell}! \;,  \quad \forall \, \balpha \in
\bN^{\ell},
\end{equation}
then we have
\begin{equation}
    \Big|\langle O^{(\ell)}, \mu_N^{(\ell)} (t) -\tilde{\mu}^{(\ell)}_t
\rangle\Big|
    \leq \, \frac{2}{\kappa_1} \,(2 C_0)^{\ell} \,  (2\kappa_t)^n +
    \frac{C_0^{\ell}}{N} \Big( 1 + \frac{3\kappa_t}{\kappa_1} \, (\ell
    +2)^2 \left( \frac{1}{1-\kappa_t}
    \right)^{\ell+3} \Big) ,
\end{equation}
where we put $\kappa_t =9\kappa_1 t (1 + 2t)(\kappa_1 +
\kappa_2)$ and assumed that $\kappa_t<1$.
\end{lemma}
\begin{proof}
{F}rom (\ref{eq:BBGKYmu}), expanding around the free evolution, we
find
\begin{equation}
\begin{split}
\mu_{N}^{(\ell)} &(t, \bxi,\bfeta) = \mu_N^{(\ell)} (0,
\bxi,\bfeta+ t\bxi) \\ &- \e^3 \sum_{1 \leq j < k \leq \ell}
\int_0^t \rd s \int \rd q \;\hat{U}(q) \; \frac{2}{\e}\sin \left(\frac{\e
q\cdot((\eta_j - \eta_k) +(t- s) (\xi_j - \xi_k))}{2}\right)\\
&\qquad \times
    \mu_N^{(\ell)}(s, \xi_1, \dots, \xi_j - q,\dots ,\xi_k + q, \dots,
\xi_n; \bfeta+(t-s) \xi) \\
    &- (1-\ell \e^3)\sum_{j=1}^{\ell} \int_0^t \rd s
    \int \rd q \;\hat{U}(q) \frac{2}{\e} \sin \left(
    \frac{\e q\cdot(\eta_j+(t-s)\xi_j)}{2}\right) \\ &\qquad \times
    \mu_N^{(\ell+1)}(s, \xi_1, \ldots ,\xi_j - q ,\dots , \xi_{\ell}, q ;
    \bfeta+(t-s)\bxi, 0).
\end{split}
\end{equation}
Next we insert this expansion in the expectation $\langle
O^{(\ell)}, \mu_N^{(\ell)} \rangle$ and we find, moving the free
evolution from the $\mu$ to the observable,
\begin{equation}\label{expa0}
\begin{split}
\langle O^{(\ell)} , \mu_N^{(\ell)}(t) \rangle = &\int \rd\bxi
\rd\bfeta \; \ov{O^{(\ell)}} (\bxi ,\bfeta - t \bxi)
\mu_N^{(\ell)} (0,\bxi,\bfeta) \\
&-  \e^3 \sum_{1 \leq j < k \leq \ell} \int_0^t \rd s  \int \rd
\bxi \rd \bfeta \int \rd q \; \hat{U}(q)  \ov{O^{(\ell)}} (\bxi ,
\bfeta - (t-s)
\bxi)\; \frac{2}{\e} \sin \left(\frac{\e q\cdot(\eta_j -
\eta_k)}{2}\right)\\
&\qquad \times
    \mu_N^{(\ell)}(s, \xi_1, \dots, \xi_j - q,\dots , \xi_k + q, \dots,
\xi_n; \bfeta) \\
    &- (1-\ell \e^3)\sum_{j=1}^{\ell} \int_0^t \rd s \int \rd\bxi
    \rd\bfeta \int \rd q \; \hat{U}(q) \ov{O^{(\ell)}} (\bxi,\bfeta - (t-s)
\bxi)
    \frac{2}{\e} \sin \left(
    \frac{\e q\cdot\eta_j}{2}\right) \\ &\qquad\times
    \mu_N^{(\ell+1)}(s, \xi_1, \ldots \xi_j - q ,\dots \xi_{\ell}, q ;
    \bfeta, 0).
\end{split}
\end{equation}
Now we define the following two operators acting on the observable
$O^{(\ell)}$:
\begin{equation}\label{eq:A}
(AO^{(\ell)}) (\bxi , \bfeta) = - \e^3 \sum_{1\leq j < k\leq
\ell} \int \rd q \; \hat{U} (q) \; \frac{2}{\e} \sin \left(\frac{\e
q\cdot(\eta_j
- \eta_k)}{2}\right)\, O^{(\ell)} (\xi_1, \dots, \xi_j +q ,\dots
,\xi_k -q ,\dots, \xi_{\ell} ; \bfeta)
\end{equation}
and
\begin{equation}\label{eq:B}
(BO^{(\ell)}) (\bxi , \xi_{\ell + 1}; \bfeta, \eta_{\ell+1}) =
-\,\sum_{j=1}^\ell \hat{U} (\xi_{\ell+1}) \delta (\eta_{\ell + 1})
\frac{2}{\e} \sin \left( \frac{\e \xi_{\ell+1}
\cdot\eta_j}{2}\right) O^{(\ell)} (\xi_1,\dots,\xi_j
+\xi_{\ell+1},\dots \xi_{\ell};\bfeta).
\end{equation}
Moreover, we denote by $(S_t O^{(\ell)}) (\bxi,\bfeta): = O^{(\ell)}
(\bxi, \bfeta - t \bxi)$ the free evolution of the observable
$O^{(\ell)}$. Equation (\ref{expa0}) can be rewritten as
\begin{equation}
\begin{split}
\langle O^{(\ell)} , &\mu_N^{(\ell)}(t) \rangle = \int \rd\bxi
\rd\bfeta \; (S_t \ov{O^{(\ell)}}) (\bxi ,\bfeta)
\mu_N^{(\ell)} (0,\bxi,\bfeta) \\
&+ \int_0^t \rd s  \int \rd\bxi \rd\bfeta \; (AS_{t-s}\ov{O^{(\ell)}})
(\bxi , \bfeta) \mu_N^{(\ell)}(s, \bxi,\bfeta) \\
    &+ (1-\ell \e^3) \int_0^t \rd s \int \rd\bxi \rd\bfeta
\rd\xi_{\ell+1}\rd\eta_{\ell+1} \, (BS_{t-s}\ov{O^{(\ell)}})
    (\bxi,\xi_{\ell+1},
    \bfeta,\eta_{\ell+1}) \mu_N^{(\ell+1)}(s, \bxi,\xi_{\ell+1},
    \bfeta, \eta_{\ell+1}),
\end{split}
\end{equation}
or, in a more compact form, as
\begin{equation}
\begin{split}
\langle O^{(\ell)} , \mu_N^{(\ell)}(t) \rangle = \; &\langle S_t
O^{(\ell)}, \mu_N^{(\ell)} (0) \rangle
+ \int_0^t \rd s  \langle AS_{t-s}O^{(\ell)} , \mu_N^{(\ell)}(s) \rangle \\
&+ (1- \ell \e^3) \int_0^t \rd s \langle BS_{t-s}O^{(\ell)},
\mu_N^{(\ell+1)}(s) \rangle ,
\end{split}
\end{equation}
where we used that the operators $A,B$ and $S_t$ commute with the
complex conjugation (note that, since $U(x)$ is  symmetric and
$\hat{U}(q)$ is real). Next we iterate this relation $n$ times. We
find
\begin{equation}\label{eq:iter}
\begin{split}
\langle O^{(\ell)}, &\mu_N^{(\ell)} (t) \rangle =  \langle S_t
O^{(\ell)}, \mu_N^{(\ell)} (0) \rangle  \\ &+ \sum_{m=1}^{n-1}
\int_0^t \rd s_1 \int_0^{s_1} \rd s_2 \dots \int_0^{s_{m-1}} \rd s_m
\langle S_{s_m} B S_{s_{m-1}-s_m} B \dots B
S_{t-s_1}O^{(\ell)}, \mu_N^{(\ell+m)}(0) \rangle \\
&+ \int_0^t \rd s_1 \int_0^{s_1} \rd s_2 \dots \int_0^{s_{n-1}} \rd s_n
\langle B S_{s_{n-1}-s_n} B \dots B S_{t-s_1}O^{(\ell)},
\mu_N^{(\ell+n)}(s_n)\rangle\\
&+ \sum_{m=1}^{n} \int_0^t \rd s_1 \int_0^{s_1} \rd s_2 \dots
\int_0^{s_{m-1}} \rd s_m \langle A S_{s_{m-1}-s_m} B \dots B
S_{t-s_1}O^{(\ell)}, \mu_N^{(\ell+ m-1)}(s_m) \rangle \\
&-\e^3 \sum_{m=1}^n (\ell + m -1) \int_0^t \rd s_1 \int_0^{s_1} \rd s_2
\dots \int_0^{s_{m-1}} \rd s_m  \langle B S_{s_{m-1} -s_{m}} B \dots
B S_{t-s_1} O^{(\ell)} , \mu_N^{(\ell+m)} (s_m) \rangle \; .
\end{split}
\end{equation}
Using (\ref{eq:sup}) we have, for any time $t$ and any observable
$O^{(k)}$,
\begin{equation}
|\langle O^{(k)} , \mu_N^{(k)}(t) \rangle| \leq \int \rd\bxi \rd\bfeta
\; |O^{(k)} (\bxi,\bfeta)|.
\end{equation}
So, in order to control the error terms on the last three lines of
(\ref{eq:iter}) we need to estimate the quantities
\begin{equation}
\begin{split}
K_{\ell,n} &:= \int
\Big|\Big(\prod_{k=1}^n\,S_{s_k}\,B\,S_{-s_k}\,S_t O^{(\ell)}
\Big) (\xi_1,\ldots,\xi_{n + \ell};\eta_1,\ldots,\eta_{n+ \ell})\,\Big|\,\rd
\bxi
\,\rd\bfeta \quad \quad \text{and} \\
M_{\ell,n} &:= \int \Big|\Big(S_{s_n}A S_{-s_n}
\prod_{k=1}^{n-1}\,S_{s_k}\,B\,S_{-s_k}\,S_t O^{(\ell)}\Big)
(\xi_1,\ldots,\xi_{n+\ell-1};\eta_1,\ldots,\eta_{n+\ell-1})\,\Big|\,\rd\bxi
\, \rd\bfeta.
\end{split}
\end{equation}
We begin by $K_{\ell,n}$. By the definition of the operator $B$
(see (\ref{eq:B})) we have, for general $m\in \bN$ and $s \in
\bR$,
\begin{multline}\label{eq:rec}
(S_{s}\,B\,S_{-s}\,O^{(m)} )(\xi_1,\dots ,\xi_{m+1}; \eta_1,
\dots, \eta_{m+1})=\frac{2}{\e}\sum_{j=1}^m \hat
U(\xi_{m+1})\,\sin\Big(\frac{\e}{2}(\eta_j - s \xi_j)
\xi_{m+1}\Big)
\\ \times \delta(\eta_{m+1}-s \xi_{m+1}) \,
O^{(m)} (\xi_1,\dots,\xi_j + \xi_{m+1},\dots,\xi_m; \eta_1,
\dots,\eta_j + s \xi_{m+1}, \dots ,\eta_m)\,.
\end{multline}
Since $|\sin x| \leq |x|$, we obtain the bound
\begin{multline}
K_{\ell,n} \le \int \rd\xi_1 \dots \rd\xi_{n+\ell} \rd\eta_1 \dots
\rd\eta_{n+\ell-1} |\hat U(\xi_{\ell+n})| \,
|\xi_{\ell+n}|\,\sum_{j=1}^{\ell +n-1}
|\eta_j - s_n \xi_j|\\
\times \Big|\Big(\prod_{k=1}^{n-1}\,S_{s_k}\,B\,S_{-s_k}\,S_t
O^{(\ell)} \Big)
(\xi_1,\ldots,\xi_{n+\ell-1};\eta_1,\ldots,\eta_{n+\ell-1})\,\Big| .
\end{multline}
Applying equation (\ref{eq:rec}) once again we find
\begin{equation}
\begin{split}
K_{\ell ,n}&\le \int \rd\xi_1 \dots \rd\xi_{n+\ell} \rd\eta_1
\dots \rd\eta_{n+\ell-1} |\hat
U(\xi_{\ell+n})||\xi_{\ell+n}|\,|\hat
U(\xi_{\ell+n-1})|\,|\xi_{\ell+n-1}|\, \delta(\eta_{n+\ell-1} -
s_{n-1} \xi_{n+\ell-1})
\\ &\times \Big(\sum_{j_1 = 1}^{\ell+n-1} |\eta_{j_1} - s_n \xi_{j_1}| \Big)
\, \sum_{j_2=1}^{n+\ell-2}
|\eta_{j_2} - s_{n-1}\xi_{j_2}|
\\
&\times \Big|\Big(\prod_{k=1}^{n-2}\,S_{s_k}\,B\,S_{-s_k} \, S_t
O^{(\ell)}\Big)(\xi_1,..,\xi_{j_2} + \xi_{n+\ell-1},.., \xi_{n+
\ell-2};\eta_1, .. ,\eta_{j_2} + s_{n-1} \xi_{n+\ell-1}, ..
\eta_{n+ \ell-2})\,\Big|.
\end{split}
\end{equation}
After shifting the variables $\xi_{j_2} \to \xi_{j_2} -
\xi_{n+\ell -1}$, $\eta_{j_2} \to \eta_{j_2} - s_{n-1}
\xi_{n+\ell-1}$ and computing the integral over $\eta_{n+\ell -1}$
(using the delta-function) we get
\begin{equation}
\begin{split}
K_{\ell,n}\le \int &\rd\xi_1 \dots \rd\xi_{n+\ell} \rd\eta_1 \dots
\rd\eta_{n+\ell-2} |\hat U(\xi_{\ell+n})||\xi_{\ell+n}|\,|\hat
U(\xi_{\ell+n-1})|\,|\xi_{\ell+n-1}|\,
\\ &\times \Big( \sum_{j_1=1}^{\ell+n-2} |\eta_{j_1} - s_n \xi_{j_1}| +
2(s_{n-1}- s_n)
|\xi_{n+\ell-1}|\Big) \,\Big( \sum_{j_2 = 1}^{n+\ell-2}
|\eta_{j_2} - s_{n-1}\xi_{j_2}| \Big) \\ &\times
\Big|\Big(\prod_{k=1}^{n-2}\,S_{s_k}\,B\,S_{-s_k} \, S_t
O^{(\ell)}\Big)(\xi_1, \dots , \xi_{n+ \ell-2};\eta_1,\dots
\eta_{n+ \ell-2})\,\Big| .
\end{split}
\end{equation}
After $n$ such iterations we arrive to the estimate
\begin{multline}\label{eq:est1}
K_{\ell,n} \le \int \rd\xi_1 \dots \rd\xi_{n+\ell} \, \rd\eta_1 \dots
\rd\eta_{\ell} \, \prod_{k=1}^{n}|\hat
U(\xi_{\ell+k})|\,|\xi_{\ell+k}|\,\\ \times \prod_{k=1}^{n}\,
\Big(\sum_{i=1}^{\ell} |\eta_i + (t - s_{k}) \xi_i| + 2
\sum_{j=\ell+1}^{\ell+k-1} (s_{j-\ell} - s_k) |\xi_j|)\Big) \,
|O^{(\ell)} (\bxi,\bfeta)| \,.
\end{multline}
Using that $|s_i - s_j| \leq t$ for all $i,j$ we get the bound
\begin{multline}
K_{\ell,n} \le \int \rd\xi_1 \dots \rd\xi_{n+\ell} \, \rd\eta_1 \dots
\rd\eta_{\ell} \, \prod_{k=1}^{n}|\hat
U(\xi_{\ell+k})|\,|\xi_{\ell+k}|\,\\ \times \prod_{k=1}^{n}\,
\Big(\sum_{i=1}^{\ell} (|\eta_i| + t |\xi_i|) + 2 t
\sum_{j=\ell+1}^{\ell+k-1} |\xi_j|)\Big) \, |O^{(\ell)}
(\bxi,\bfeta)| \,.
\end{multline}
Let us use the notation
\begin{equation}
x_1:=\sum_{i=1}^{\ell}( |\eta_i| + t |\xi_i|)\,;\quad x_{j}:= 2 t
|\xi_{\ell+j-1}|\ \;\;\; {\rm for }\ j=2,\dots , n\,.
\end{equation}
The integrand on the right hand side of equation (\ref{eq:est1}) is
dominated by
\begin{equation}
\Big( \prod_{k=1}^{n}|\hat U(\xi_{\ell+k})|\,|\xi_{\ell+k}|
\Big)\cdot
|O^{(\ell)}(\xi_1,\ldots,\xi_{\ell};\eta_1,\ldots,\eta_{\ell})|
\,\cdot\,\prod_{k=1}^{n}\,\sum_{j=1}^k
x_j\,,
\end{equation}
which, in turn, is bounded by
\begin{equation}
\Big( \prod_{k=1}^{n}|\hat U(\xi_{k+1})|\,|\xi_{k+1}|\Big)\cdot
|O^{(\ell)}(\xi_1,\ldots,\xi_{\ell};\eta_1,\ldots,\eta_{\ell})|\,\cdot\,\Big(\sum_{j=1}^n
\frac{n-j+1}{n} x_j \Big)^n\,,
\end{equation}
where we estimated the product by its arithmetic mean in power
$n$. Next we use the binomial expansion
$$
\Big(\sum_{j=1}^n
\frac{n-j+1}{n}x_j\Big)^n = n! \sum_{\alpha_1+\ldots+\alpha_n=n}
\prod_{j=1}^n\frac{\Big(\frac{n-j+1}{n}x_j\Big)^{\alpha_j}}{\alpha_j!}
\,,$$
and we note that, because of the assumption (\ref{eq:ass2}),
we have
\begin{multline}\label{eq:bnd1}
\int |O^{(\ell)}(\xi_1,\ldots,\xi_{\ell};\eta_1,\ldots,\eta_{\ell})|\,
\left(\sum_{i=1}^{\ell} |\eta_i| + t |\xi_i|\right)^{\alpha
}\,\rd\xi\,\rd\eta\\
\le (1 + t)^{\alpha} \, \sum_{\alpha_1+\ldots+\alpha_{\ell}=\alpha}
\frac{\alpha!}{\prod\alpha_i!}\int
|O^{(\ell)}(\xi_1,\ldots,\xi_{\ell};\eta_1,\ldots,\eta_{\ell})|\,
\prod_{i=1}^{\ell} \,(|\eta_i| + |\xi_i|)^{\alpha_i}\,\rd\xi\,\rd\eta\\
\le C_0^{\ell} (1 + t)^{\alpha} \, \kappa_2^{\alpha} \, \alpha!
\sum_{\alpha_1+ \ldots+\alpha_{\ell} =\alpha} 1 \leq C_0^{\ell}\, (1 +
t)^{\alpha} \, \kappa_2^{\alpha} \,
\frac{(\alpha+\ell)!}{\ell!}\,.
\end{multline}
This, together with the assumption (\ref{eq:ass1}), implies that
\begin{equation}\label{eq:Kln}
\begin{split}
K_{\ell,n}&\le  C_0^{\ell} \, n! \sum_{\alpha_1+\ldots+\alpha_n=n}
\kappa_2^{\alpha_1} (1 + t)^{\alpha_1} \, \kappa_1^{2n-\alpha_1-1}
(2t)^{n-\alpha_1} \frac{(\alpha_1+\ell)!}{\alpha_1!\ell!}
\prod_{j=2}^n \, (\alpha_j +1) \, \Big(\frac{n-j+1}{n}\Big)^{\alpha_j}\\
&\le C_0^{\ell} \,\kappa_1^{n-1} ((1+t) \kappa_1 + 2t
\kappa_2)^{n} \frac{(n+\ell)!}{\ell!} \prod_{j=2}^n
\Big(\frac{1}{1-\frac{n-j+1}{n}}\Big)^2 \\ &\leq C_0^{\ell}
\frac{(n+\ell)!}{\ell!} \,\kappa_1^{n-1} (\kappa_1 + \kappa_2)^{n}
(1+ 2t)^n \Big(\frac{n^{n-1}}{(n-1)!}\Big)^2 \leq n! {n+\ell
\choose \ell} \, C_0^{\ell} \kappa_1^{-1} \, [9 \kappa_1 (\kappa_1 +
\kappa_2)(1+
2t)]^n\; .
\end{split}
\end{equation}

Analogously we can bound $M_{\ell,n}$. Using the definition of $A$
we find
\begin{multline}
M_{\ell,n} \le  \e^3 \int \rd q \; |\hat
U(q)|\,|q|\,\sum_{j<k} |(\eta_j - s_n\xi_j) - (\eta_k - s_n\xi_k)|\\
\cdot \Big|\Big(\prod_{r=1}^{n-1}\,S_{s_r}\,B\,S_{-s_r}\,S_{t}
O^{(\ell)}\Big)
(\xi_1,\ldots,\xi_{\ell+n-1};\eta_1,\ldots,\eta_{\ell+n-1})\,\Big|\,\rd\bxi\,\rd\bfeta\,,
\end{multline}
and since $$\sum_{j<k} |\eta_j-s_n\xi_j - (\eta_k-s_n\xi_k)|\le
(\ell+n-2)\sum_{j=1}^{\ell+n-1} |\eta_j-s_n\xi_j|\,,$$ we get the
bound
\begin{equation}\label{eq:Mln}
M_{\ell,n} \le \e^3 (\ell+n-2) K_{\ell,n}\,.
\end{equation}

Inserting (\ref{eq:Kln}) and (\ref{eq:Mln}) in (\ref{eq:iter}) and
performing the integration over the $s$ variables, we find
\begin{multline}\label{err1}
\Big| \langle O^{(\ell)} , \mu_N^{(\ell)} (t) \rangle - \Big\{
\langle S_t O^{(\ell)}, \mu_N^{(\ell)} (0) \rangle  + \\  +
\sum_{m=1}^{n-1} \int_0^t \rd s_1 \int_0^{s_1} \rd s_2 \dots
\int_0^{s_{m-1}} \rd s_m \langle S_{s_m} B S_{s_{m-1}-s_m} B \dots B
S_{t-s_1}O^{(\ell)}, \mu_N^{(\ell+m)}(0) \rangle \Big\} \Big| \\
\leq {n+\ell \choose \ell} \, C_0^{\ell} \kappa_1^{-1} \,
\kappa_t^n + 2 \kappa_1^{-1} C_0^{\ell} \e^3 \sum_{m=1}^{n}
(\ell+m) {m+\ell \choose m} \kappa_t^m \; ,
\end{multline}
where we introduced
$\kappa_t: = 9 t\kappa_1 (\kappa_1 + \kappa_2)(1+
2t)$. Next we want to compare $\langle O^{(\ell)} , \mu_N^{(\ell)}
(t)\rangle$ with $\langle O^{(\ell)} , \tilde{\mu}_t^{(\ell)}
\rangle$, where $\tilde{\mu}_t^{(\ell)}$ was defined in
(\ref{eq:Hprod}). Using that $\mu_t (\xi, \eta)$ is a solution of
the 1-particle Hartree Equation (\ref{eq:hartree}) we find that
$\tilde{\mu}_t^{(\ell)}$ satisfies the following hierarchy of
equation:
\begin{equation}
\begin{split}
\partial_t \tilde{\mu}_t^{(\ell)}(\bxi, \bfeta)
    = &\, \bxi \cdot \nabla_{\bfeta} \tilde{\mu}_t^{(\ell)}(\bxi, \bfeta) \\
    &- \sum_{j=1}^{\ell}
    \int \rd q \hat{U}(q) \frac{2}{\e} \sin \left(
    \frac{\e q\cdot\eta_j}{2}\right)
    \tilde{\mu}_t^{(\ell+1)}( \xi_1, \ldots,\xi_j - q ,\dots ,\xi_n, q ;
    \bfeta, 0) .
\end{split}
\end{equation}
One can then expand the expectation $\langle O^{(\ell)},
\tilde{\mu}_t^{\ell}\rangle$ in a series, exactly as we did for
$\langle O^{(\ell)},  \mu_N^{\ell} (t)\rangle$. Clearly one finds
\begin{equation}
\begin{split}
\langle O^{(\ell)} , \tilde{\mu}_t^{(\ell)} \rangle =  &\, \langle
S_t O^{(\ell)}, \tilde{\mu}_{0}^{(\ell)} \rangle  \\ &+
\sum_{m=1}^{n-1} \int_0^t \rd s_1 \int_0^{s_1} \rd s_2 \dots
\int_0^{s_{m-1}} \rd s_m \langle S_{s_m} B S_{s_{m-1}-s_m} B \dots B
S_{t-s_1}O^{(\ell)}, \tilde{\mu}_0^{(\ell+m)} \rangle \\
&+ \int_0^t \rd s_1 \int_0^{s_1} \rd s_2 \dots \int_0^{s_{n-1}} \rd s_n
\langle B S_{s_{n-1}-s_n} B \dots B S_{t-s_1}O^{(\ell)},
\tilde{\mu}_{s_n}^{(\ell+m)}\rangle.
\end{split}
\end{equation}
The error term on the last line can be bounded as before (equation
(\ref{eq:sup}) holds with $\mu_N^{(k)}$ replaced by
$\tilde{\mu}^{(k)}$ as well). We have
\begin{multline}\label{err2}
\Big|\langle O^{(\ell)} , \tilde{\mu}_t^{(\ell)} \rangle  - \Big\{
\langle S_t O^{(\ell)}, \tilde{\mu}_{0}^{(\ell)} \rangle + \\
+ \sum_{m=1}^{n-1} \int_0^t \rd s_1 \int_0^{s_1} \rd s_2 \dots
\int_0^{s_{m-1}} \rd s_m \langle S_{s_m} B S_{s_{m-1}-s_m} B \dots B
S_{t-s_1}O^{(\ell)}, \tilde{\mu}_0^{(\ell+m)} \rangle  \Big\}\Big|
\\ \leq {n+\ell \choose \ell} \, C_0^{\ell} \kappa_1^{-1} \, \kappa_t^n \, .
\end{multline}
Combining (\ref{err1}) and (\ref{err2}) we find
\begin{equation}
\begin{split}
\Big| \langle &O^{(\ell)} , (\mu_N^{(\ell)} (t) -
\tilde{\mu}_t^{(\ell)}) \rangle \Big| \leq 2 {n+\ell \choose \ell}
\, C_0^{\ell} \kappa_1^{-1} \, \kappa_t^n + 2 \e^3 C_0^{\ell}
\kappa_1^{-1} \sum_{m=1}^{n} (\ell+m) {m+\ell \choose m} \,
\kappa_t^m
\\ &+\Big| \langle
S_t O^{(\ell)} , (\mu_N^{(\ell)} (0) - \tilde{\mu}_0^{(\ell)}) \rangle\Big|
\\
&+ \sum_{m=1}^{n-1}\int_0^t \rd s_1 \int_0^{s_1} \rd s_2 \dots
\int_0^{s_{m-1}} \rd s_m \Big|\langle S_{s_m} B S_{s_{m-1}-s_m} B
\dots B S_{t-s_1}O^{(\ell)}, (\mu_N^{(\ell+m)} (0) -
\tilde{\mu}_0^{(\ell+m)}) \rangle \Big|\, .
\end{split}
\end{equation}
Using the assumption (\ref{eq:ass}) and equation (\ref{eq:Kln}) to
bound $\|S_{s_m} B S_{s_{m-1}-s_m} B \dots B S_{t-s_1}O^{(\ell)}
\|_0$ we find
\begin{equation}
\Big| \langle O^{(\ell)} , \mu_N^{\ell} (t) - \tilde{\mu}_t^{\ell}
\rangle \Big| \leq 2 {n+\ell \choose \ell} \, C_0^{\ell}
\kappa_1^{-1} \, \kappa_t^n + C_0^{\ell} \e^3 + 3 \e^3 C_0^{\ell}
\kappa_1^{-1} \sum_{m=1}^{n} (\ell+m) {m+\ell \choose m}
\kappa_t^m .
\end{equation}
Using that
\[ {n+ \ell \choose n} \le 2^{n+\ell} ,\] and that
\begin{equation}
\sum_{m=1}^{\infty} (\ell + m) {m+\ell \choose m} \kappa_t^m \leq
(\ell +2)^2 \, \kappa_t \sum_{m=0}^{\infty} {m + \ell + 2 \choose
m} \kappa_t^m = (\ell +2)^2 \, \kappa_t \, \left( \frac{1}{1-
\kappa_t }\right)^{\ell + 3},
\end{equation}
the claim of Lemma \ref{lm:mu} follows.
\end{proof}

In order to apply this lemma to prove Theorem \ref{thm}, we need
to estimate the $\balpha$-norm of some product of Gaussian
functions in the $\bxi$- and in the $\bfeta$-space. This is the
aim of the following lemma.
\begin{lemma}\label{lm:gauss}
For $\bxi=(\xi_1, \ldots, \xi_\ell)$, $\bfeta=(\eta_1, \ldots ,
\eta_\ell)$ we set \be F^{(\ell)}_{\delta_1, \delta_2}
(\bxi,\bfeta):= \, e^{-\frac{\delta_1^2 \bxi^2}{4}} \,
e^{-\frac{\delta_2^2 \bfeta^2}{4}}. \ee Then there exist universal
constants  $C_1$ and $C_2$ such that  for arbitrary $\kappa >0$
\begin{equation}\label{eq:claim}
\| F^{(\ell)}_{\delta_1,\delta_2} \|_{\balpha} \leq \Big(
\frac{C_1}{\delta_1^3 \delta_2^3}\Big)^{\ell} \;
C_2^{\ell/(\delta^{2} \kappa^{2})} \; \kappa^{|\balpha|} \alpha_1!
\dots \alpha_{\ell}! \; ,
\end{equation}
where $\delta^{-1} := \delta_1^{-1} + \delta_2^{-1}$, and
$|\balpha| = \alpha_1 + \dots + \alpha_{\ell}$.
\end{lemma}
\begin{proof}
We have
\begin{equation}\label{eq:b1}
\begin{split}
\| F^{(\ell)}_{\delta_1 , \delta_2} \|_{\balpha} = &\, \int \rd
\bxi \rd \bfeta \; |F^{(\ell)}_{\delta_1 , \delta_2} \,
(\bxi,\bfeta)| \, \prod_{j=1}^{\ell} (|\xi_j| +
|\mu_j|)^{\alpha_j} = \int \rd \bxi \rd \bfeta \;
e^{-\frac{\delta_1^2 \bxi^2}{4}} \, e^{-\frac{\delta_2^2
\bfeta^2}{4}} \, \prod_{j=1}^{\ell}
(|\xi_j| + | \eta_j|)^{\alpha_j} \\
\leq &\, \prod_{j=1}^{\ell} 2^{\alpha_j} \Big\{ \int \rd \xi_j
e^{-\frac{\delta_1^2 \xi_j^2}{4}} |\xi_j|^{\alpha_j} \, \int \rd
\eta_j e^{-\frac{\delta_2^2 \eta_j^2}{4}} + \int \rd \xi_j
e^{-\frac{\delta_1^2 \xi_j^2}{4}} \, \int \rd \eta_j
e^{-\frac{\delta_2^2 \eta_j^2}{4}} |\eta_j|^{\alpha_j} \Big \} \\
= &\, \left(\frac{C_1}{\delta_1^3 \delta_2^3}\right)^{\ell} \,
\prod_{j=1}^{\ell} \left(\frac{4}{\delta}\right)^{\alpha_j} \,
\Gamma \left(\frac{\alpha_j +3}{2}\right),
\end{split}
\end{equation}
for a universal constant $C_1$.
Here we put $\delta = (\delta_1^{-1} +
\delta_2^{-1})^{-1}$. Simple estimate shows that
\be
  \Gamma \left(\frac{\alpha_j +3}{2}\right) \leq
  \frac{D_1^{\alpha_j +1}}{\alpha_j^{\alpha_j/ 2}} \; \alpha_j ! \; ,
\ee thus
\begin{equation}\label{eq:F1}
\| F^{(\ell)}_{\delta_1, \delta_2} \|_{\balpha} \leq
\left(\frac{C_1}{\delta_1^3 \delta_2^3}\right)^{\ell} \,
\prod_{j=1}^{\ell} \left(\frac{D_2}{\delta^2
\alpha_j}\right)^{\alpha_j/2} \alpha_j ! \; ,
\end{equation}
where $C_1, D_2$ are universal constants. Elementary calculation
shows that \be \label{eq:F2} \left(\frac{D_2}{\delta^2
\alpha_j}\right)^{\alpha_j/2} \leq C_2^{1/(\delta^2\kappa^2)}
\kappa^{\alpha_j} \ee for a sufficiently large universal constant
$C_2$.
\end{proof}

We are now ready to proceed with the proof of our main result,
Theorem \ref{thm}.

\begin{proof}[Proof of Theorem \ref{thm}]
First we note that the assumption (\ref{eq:ass3}) is equivalent to
the assumption (\ref{eq:ass}) in Lemma \ref{lm:mu} after taking Fourier
transform.
On the other
hand, with the notation $\delta W^{(\ell)} (t) := W_N^{(\ell)} (t)
- \widetilde{W}^{(\ell)} (t)$ we have
\begin{multline}\label{eq:Wmu}
\left( H^{\delta_1 , \delta_2}_{\ell , N} - \widetilde{H}^{
\delta_1 , \delta_2}_{\ell, N} \right) (t, \bx,\bv) = \int \rd
\bx' \rd \bv' \; G_{\delta_1}^{(\ell)} (\bx-\bx')
G_{\delta_2}^{(\ell)} (\bv - \bv') \delta W^{(\ell)}
(t,\bx',\bv')\\
=\left(\frac{1}{2\pi}\right)^{6 \ell} \int \rd \bxi \rd \bfeta \,
e^{i(\bx\cdot \bxi+\bv\cdot\bfeta)} e^{-\frac{\delta_1^2
\bxi^2}{4}} \, e^{-\frac{\delta_2^2 \bfeta^2}{4}} \delta
\mu^{(\ell)} (t ,\bxi ,\bfeta) \, .
\end{multline}
In the following we use the notation
\[\tilde{F}^{(\ell)}_{\delta_1, \delta_2} (\bxi,\bfeta):=
\left(\frac{1}{2\pi}\right)^{6\ell} \, e^{i(\bx\cdot\bxi+\bv\cdot\bfeta)}
\, \exp{\Big( -\frac{\delta_1^2 \bxi^2}{4} \,
-\frac{\delta_2^2 \bfeta^2}{4}\Big)}.\]
{F}rom Lemma \ref{lm:gauss}
we find, for an arbitrary $\kappa_2 >0$,
\begin{equation}
\| \tilde{F}^{(\ell)}_{\delta_1 , \delta_2} \|_{\balpha} \leq
\Big( \frac{C_1}{2\pi\delta_1^3 \delta_2^3}\Big)^{\ell} \;
C_2^{\ell/( \delta^2 \kappa_2^2)} \; \kappa_2^{|\balpha|}
\alpha_1! \dots \alpha_{\ell}! \;
\end{equation}
where the constants $C_1$ and $C_2$ are universal. {F}rom Lemma
\ref{lm:mu} and from Eq. (\ref{eq:Wmu}) it follows that
\begin{equation*}
\begin{split}
\Big| \left( H^{\delta_1 , \delta_2}_{\ell, N} -
\widetilde{H}^{\delta_1 , \delta_2}_{\ell , N} \right) (t,
\bx,\bv) \Big| \leq \; &2 \kappa_1^{-1}\,
\Big(\frac{C_1}{\pi\delta_1^3 \delta_2^3}\Big)^{\ell}  C_2^{\ell/(\delta^2
\kappa_2^2)}
\; ( 2\kappa_t )^{n}
\\ &+ \frac{1}{N} \, \Big(\frac{C_1}{2\pi\delta_1^3
\delta_2^3}\Big)^{\ell} \, C_2^{\ell / (\delta^2 \kappa_2^2)}
\,\Big( 1 + \frac{3\kappa_t}{\kappa_1}\, (\ell+2)^2 \,
\left(\frac{1}{1 -\kappa_t}\right)^{\ell + 3}\Big)
\end{split}
\end{equation*} for any $\kappa_2 > 0$ and $n \leq 2\log N
-\ell$. Here, as in Lemma \ref{lm:mu}, we use the notation
$\kappa_t = 9 \kappa_1 (\kappa_1 + \kappa_2) t (1+2t)$. Since $t <
\sfrac{1}{4} (\sqrt{1+ 1/ (7\kappa_1^2)} -1)$, we can fix $\kappa_2
>0$ such that \( 2\kappa_t \leq
e^{-1}\). Then choosing $n= \log N$ we find
\begin{equation}
\Big| \left( H^{\delta_1 , \delta_2}_{\ell ,N} -
\widetilde{H}^{\delta_1, \delta_2}_{\ell , N} \right) (t, \bx,\bv)
\Big| \leq \frac{C_{\ell, \delta_1, \delta_2}}{N},
\end{equation}
where $C_{\ell,\delta_1,\delta_2}$ is independent of $N$. Thus,
for any fixed $\ell, \delta_1,\delta_2$ we get
\begin{equation}
\limsup_{N\to\infty} \sup_{\bx,\bv \in \bR^{3\ell}} \Big| \left(
H^{ \delta_1  , \delta_2 }_{\ell , N} - \widetilde{H}^{ \delta_1 ,
\delta_2 }_{\ell , N} \right) (t, \bx,\bv) \Big| \cdot N  \leq
C_{\ell,\delta_1,\delta_2}.
\end{equation}
\end{proof}

\appendix

\section{Proof of Lemma \ref{lemma:lt} and \ref{lemma:mourre}}

{\it Proof of Lemma \ref{lemma:lt}.} We can restrict ourselves to pure states.
Let $\Psi$ be a normalized fermionic wavefunction. For any $X\in \bR^3$ define
$$
      \Psi_X(y_1, \ldots , y_{N-1}): = \Psi\big(y_1 + \bar X, y_2 + \bar X, \ldots, y_{N-1}+\bar X, \bar X-(y_1+\ldots +y_{N-1})\big)\; ,
$$
where $\bar X:= X/N$.
Clearly $\Psi_X$ is antisymmetric and $\int \|\Psi_X\|^2 \rd X =1$.
By the Lieb-Thirring inequality in the Fourier space (\ref{eq:LT})
\be
      \int \varrho_X(v)^{5/3} \, \rd v\leq (const.) \| \Psi_X\|^{4/3}
      \Big\langle \Psi_X, \Big( \sum_{j=1}^{N-1} y_j^2\Big)
      \Psi_X\Big\rangle\; ,
\label{eq:1}
\ee
where $\varrho_X:=\varrho_{\Psi_X}$ is the momentum distribution of the one-particle
 marginal of $\Psi_X$
with the normalization $\int \varrho_X = (N-1)\| \Psi_X\|^2$ (see the proof
of Lemma \ref{lemma:wigner}).
Simple calculation shows that
\be
    \int \rd X  \, \Big\langle \Psi_X, \Big( \sum_{j=1}^{N-1} y_j^2\Big)
      \Psi_X\Big\rangle = \frac{N-1}{N}   \Big\langle \Psi, \sum_{j=1}^N ( x_j - \bar X)^2
      \; \Psi \Big\rangle \; .
\label{eq:2} \ee For an arbitrary $\rho (v) \in L^1 (\bR^3) \cap
L^{5/3} (\bR^3)$ we have
\begin{equation}
\begin{split}
\int_{|v| \geq \ell} \rd v \, \rho (v) &\leq \frac{1}{\ell^2}
\int_{|v|\geq \ell} \rd v \, v^2 \rho (v) \leq \frac{1}{\ell^2}
\int \rd v \, v^2 \rho (v) ,\\
\int_{|v| \leq \ell} \rd v \, \rho (v) &\leq \ell^3 \, \left(
\frac{1}{\ell^3} \int_{|v| \leq \ell} \rd v \, \rho^{5/3} (v)
\right)^{3/5} \leq \ell^{6/5} \| \rho \|_{5/3}.
\end{split}
\end{equation}
This implies that
\begin{equation}
\int \rd v \, \rho (v) \leq \frac{1}{\ell^2} \int \rd v \, v^2
\rho (v) + \ell^{6/5} \| \rho \|_{5/3}.
\end{equation}
Optimizing with respect to $\ell$ we easily obtain $\int
v^2\varrho(v)\rd v \ge (const.) \|\varrho\|_1^{8/3} /
\|\varrho\|_{5/3}^{5/3}$ with a positive constant. Applying this
inequality for $\varrho_X$, using (\ref{eq:1}) and the
normalization $\| \varrho_X \|_1= (N-1)\| \Psi_X\|^2$, we have
$$
      \int v^2 \varrho_X(v) \, \rd v \, \ge
      \frac{(const.) (N-1)^{8/3} \|\Psi_X\|^4}{  \Big\langle \Psi_X, \Big( \sum y_j^2\Big)
      \Psi_X\Big\rangle} \; .
$$
Integrating $X$, using a Schwarz inequality and (\ref{eq:2}) we obtain
$$
     \int \!\!\int v^2 \varrho_X(v) \, \rd v  \,\rd X \ge (const.) N^{8/3}
    \frac{ \big(\int \|\Psi_X\|^2 \, \rd X\big)^2}{\int \big\langle \Psi_X, \Big( \sum y_j^2\Big)
      \Psi_X\big\rangle \, \rd X} \ge  (const.) N^{5/3}
$$
if $N\ge 2$.
Finally we conclude by the identity
$$
    \int \!\! \int v^2 \; \varrho_X(v)  \,\rd v \,\rd X
    = \int \rd X \, \Big\langle \Psi_X, \Big( \sum_{j=1}^{N-1} (p_j-p_N)^2 \Big)\Psi_X\Big\rangle
    = (N-1)\langle \Psi, (p_1-p_2)^2 \Psi\rangle\; ,
$$
where we again used the symmetry of $\Psi$.  $\;\;\Box$

\bigskip\bigskip

{\it Proof of Lemma \ref{lemma:mourre}}.  First we want
to prove that $v_t$ remains of order $N^{1/3}$ for all finite
times. To this end we compute
\begin{equation*}
\begin{split}
[iH_{N,\alpha} , (p_1 -p_2)^2] = \; -\frac{1}{N} \sum_{m\geq 3}
\Big( &(p_1 -p_2)\cdot (\nabla U (x_1 -x_m) - \nabla U (x_2 -x_m))
\\
&+(\nabla U (x_1 -x_m) - \nabla U (x_2 -x_m))\cdot (p_1 -p_2) \Big) \\
-\frac{2}{N} \Big( (p_1-&p_2)\cdot \nabla U (x_1 -x_2) + \nabla U
(x_1 -x_2) \cdot (p_1 -p_2) \Big) ,
\end{split}
\end{equation*}
which implies, using $C=\| \nabla U \|_\infty$, and
applying the Schwarz inequality, that
\begin{equation*}
|\partial_t v_t^2 | = \big| \tr \; \big(\gamma_{N,t}[iH_{N,\alpha} , (p_1 -p_2)^2]\big)\big|
  \leq 8 C \, v_t.
\end{equation*}
Integrating the last equation we obtain
\begin{equation}\label{eq:vt}
 v_0 - 4Ct \leq v_t \leq v_0 + 4Ct \;
\end{equation}
for all $t>0$.
Next we derive an upper bound for the quantity
$u_t$. Here we use
\begin{equation*}
[iH_{N,\alpha} , (x_1 -x_2)^2] = 2\alpha \Big( (p_1 -p_2)\cdot
(x_1 -x_2) + (x_1 -x_2)\cdot (p_1 -p_2) \Big),
\end{equation*}
and from (\ref{eq:vt}) we find that
\begin{equation}\label{eq:derbound}
|\partial_t u_t^2 | \leq 4 \alpha \, v_t \, u_t \leq 4
\, \alpha \, (v_0+4Ct) \, u_t,
\end{equation}
hence, for $t\leq v_0 /8C$,
\begin{equation}\label{eq:ut}
u_t \leq u_0 + 3 \, \alpha \, v_0 \, t .
\end{equation}
Finally we want to estimate the
quantity $u_t$ from below. To this end we compute the second
derivative of $u_t$ using that
\begin{equation*}
\begin{split}
[iH_{N,\alpha}, [iH_{N,\alpha} , (x_1 -x_2)^2]] = \; &8 \alpha^2
(p_1 -p_2)^2 - \frac{4\alpha}{N} \sum_{m\geq 3} (\nabla U (x_1
-x_m) -\nabla U (x_2 -x_m)) \cdot (x_1 - x_2) \\
&-\frac{8\alpha}{N} \nabla U (x_1 -x_2)\cdot (x_1 -x_2).
\end{split}
\end{equation*}
Applying the Schwarz inequality, using  $C=\| \nabla U \|_\infty$
 and equations (\ref{eq:vt}),  (\ref{eq:ut}), we find
\begin{equation*}
\begin{split}
\partial_t^2 u_t^2 &\geq  8 \, \alpha^2 \, v_t^2 - 8C \, \alpha \, u_t \\
&\geq 2 \alpha^2 \, v_0^2 - 8C \, \alpha  \, (u_0 + 3 \, \alpha \,
v_0 t)\; ,
\end{split}
\end{equation*}
for $t \leq v_0 / 8C$. Integrating this equation twice with the
help of (\ref{eq:derbound}), one easily finds \eqref{eq:lowu}.
$\;\;\Box$

\bigskip\bigskip

\noindent
{\it Authors' email addresses:}

\noindent
Alexander Elgart: elgart@math.stanford.edu

\noindent
L\'aszl\'o Erd{\H o}s: lerdos@mathematik.uni-muenchen.de

\noindent
Benjamin Schlein: schlein@math.stanford.edu

\noindent
Horng-Tzer Yau: yau@math.stanford.edu

\end{document}